\documentclass[pre,amsmath,amssymb,groupedaddress]{revtex4-2}
\usepackage{url}
\usepackage{setspace} 
\usepackage{lmodern}
\usepackage{graphicx}
\usepackage{gensymb}

\usepackage{amsmath}
\usepackage[colorlinks=true, citecolor=blue, linkcolor=blue, urlcolor=red]{hyperref}

\usepackage{natbib}
\newcommand{\RomanNumeralCaps}[1]
\linenumbers
\newcommand\Rey{{\mathrm{Re}}}
\newcommand\Pe{{\mathrm{Pe}}}

\begin{document}

\title{Viscosity measurements of glycerol in a parallel-plate rheometer exposed to atmosphere}

\author{Jesse T. Ault}
\email{jesse\_ault@brown.edu}
\affiliation{Center for Fluid Mechanics, Brown University, Providence, Rhode Island 02912, USA}
\author{Sangwoo Shin}
\affiliation{Department of Mechanical and Aerospace Engineering, University at Buffalo, The State University of New York, Buffalo, NY 14260, USA}
\author{Allan Garcia}
\affiliation{Department of Mechanical and Aerospace Engineering, Princeton University, Princeton, NJ 08544, USA}
\author{Antonio Perazzo}
\affiliation{Department of Mechanical and Aerospace Engineering, Princeton University, Princeton, NJ 08544, USA}
\author{Howard A. Stone}
\affiliation{Department of Mechanical and Aerospace Engineering, Princeton University, Princeton, NJ 08544, USA}

\begin{abstract}
Glycerol is a hygroscopic fluid that spontaneously absorbs water vapor from the atmosphere. For applications involving glycerol, care must be taken to avoid exposure to humidity, since its viscosity decreases quickly as water is absorbed. We report experimental measurements of the viscosity of glycerol in a parallel-plate rheometer where the outer interface is exposed to atmosphere. The measurements decrease with time as water is absorbed from the atmosphere and transported throughout the glycerol via diffusion and advection. Measured viscosities drop faster at higher relative humidities, confirming the role of hygroscopicity on the transient viscosities. The rate of viscosity decrease shows a non-monotonic relationship with the rheometer gap height. This behavior is explained by considering the transition from diffusion-dominated transport in the narrow gap regime to the large gap regime where transport is dominated by inertia-driven secondary flows. Numerical simulations of the water absorption and transport confirm this non-monotonic behavior. The experimental viscosity measurements show unexpectedly fast decreases at very small gap heights, violating the parallel-plate, axisymmetric model. We propose that this drop-off may be due to misalignment in the rheometer that becomes non-negligible for small gaps. Theoretical considerations show that secondary flows in a misaligned rheometer dominate the typical secondary inertial flows in parallel-plate rheometers at small gaps. Finally, simulations in a misaligned parallel-plate system demonstrate the same sharp drop-off in viscosity measurements at small gap heights. This modeling can be used to estimate the gap height where misalignment effects dominate the transient glycerol viscosity measurements.
\end{abstract}

\maketitle

\section{Introduction}
A rotational rheometer is a device in which one component rotates relative to another in order to induce a shear on the fluid placed in between the two components. Through a characterization of the torques and forces that result as a function of rotation rate, rheological properties of the fluids can be measured. In the last several decades, rheometry has emerged as an essential tool for studying the fluid dynamics of complex fluids for measuring properties ranging from the dynamic viscosity of Newtonian fluids to the viscoelastic responses of non-Newtonian fluids \citep{malkin2017rheology, coussot2005rheometry}. Owing to their precision and versatility, rheometers have been utilized to investigate the rheology and mechanics of a wide range of viscous and viscoelastic fluids such as polymer melts, gels, suspensions, cells, bacterial biofilms, lipid vesicle solutions, food products, cosmetics, pharmaceuticals, and many others \citep{dakhil2018measuring, yan2018bacterial, shin2015flow, gallegos1999rheology, kavehpour2004tribo, clasen2010bridging, dhinojwala1997micron}. Rheometry has also been used for evaluating drag reduction, since these techniques can also be used to characterize slip lengths such as those generated by nanostructured surfaces \citep{choi2006large, bocquet2006comment, srinivasan2013drag}. Among various configurations, parallel-plate rheometry is a system that is commonly used for highly viscous and viscoelastic fluids due to the ability to carefully control the gap height and the spatially uniform confinement in the system. From a fluid dynamics perspective, a parallel-plate rheometer generates a shear-driven fluid velocity that is primarily in the azimuthal direction. In the limit of slow rotation speed or narrow gap heights, the velocity profile in such a system is simply $(u_r,u_\theta,u_z)=(0,\Omega r z/h_0,0)$, where $\Omega$ is the angular velocity of the upper plate, $h_0$ is the gap height between the plates, and $(r,\theta,z)$ are the cylindrical coordinates with the origin located at the center of the bottom plate (see e.g., \citet{middleman1968flow,bird1987dynamics} and others).

In addition to the primary azimuthal flow, secondary recirculating flows exist due to inertial effects for any finite angular rotation speed \citep{greensmith1953hydrodynamics, ewoldt2015experimental}. The first experimental evidence of these secondary flows was achieved by \citet{garner1950thermodynamics} in a study of the rheological properties of a hydrocarbon-type micellar system. The secondary flow profile in a parallel-plate rheometer was found by \citeauthor{savins1970radial} more than half a century ago and has been well-described by various authors (see e.g., \citet{savins1970radial,denn1980process}), where the radial velocity (except near the turning regions) is given as
\begin{equation}\label{eq:savins}
u_r(r,z) = -\frac{\rho \Omega^2 h_0^2 r }{12\mu}\left[ \frac{4}{5}\left(\frac{z}{h_0}\right) - \frac{9}{5}\left(\frac{z}{h_0}\right)^2 + \left(\frac{z}{h_0}\right)^4 \right], 
\end{equation}
where $\rho$ is the fluid density and $\mu_0$ is the fluid viscosity. Notably, the radial flow becomes increasingly important as the gap height or rotation speed increase (i.e., $u_r/u_\theta \sim \Omega h_0^2/\nu$).

The secondary fluid dynamics can play a surprisingly significant role in a rheometer both by altering the torque measurements and by triggering instabilities or driving fluid mixing, especially in the case of non-homogeneous fluids. For example, \citet{jacobi2015stratified} studied the effect of radial flow on the viscosity measurements in parallel-plate and cone-and-plate settings when the target fluid is stratified with an immiscible fluid. In this case, the authors found that the radial flow can distort the fluid interface, leading to drastically different torque measurements and even fluid dewetting. Another situation in which the secondary recirculation plays a key role is in the case of an initially homogeneous fluid that becomes non-homogeneous during the measurement procedure due to mass transfer that occurs at the exposed outer edge, such as by solute sorption or solvent evaporation/condensation. This is particularly important for parallel-plate rheometry since inhomogeneities alter the stress profiles, such that small changes in the fluid composition at the outer edge can have significant impacts on the torque measurement. This secondary flow is expected to play a significant role in the viscosity measurements of glycerol over time, since any water absorbed at the outer edge of the rheometer can be transported radially inwards by the secondary flow, redistributing the relatively lower viscosity glycerol (where the water fraction is higher). Since the torque in a parallel-plate rheometer is primarily generated near the outer edge of the system, this redistribution of the lower-viscosity fluid must have direct consequences on the measured torque value.

\begin{figure*}
\centering\includegraphics[width=\textwidth]{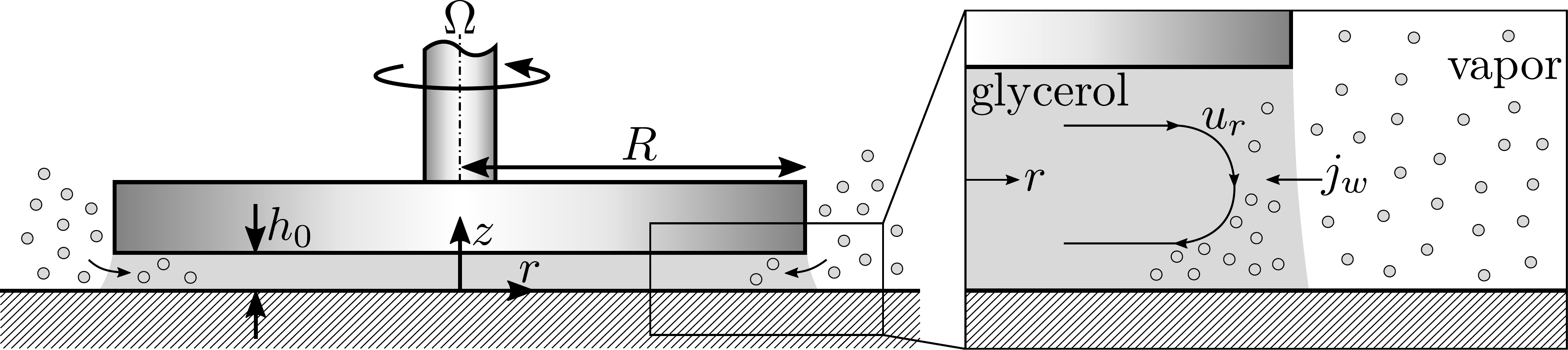}
\caption{Problem setup. We measure the viscosity of glycerol in a parallel-plate rheometer exposed to atmosphere at the outer fluid interface. Due to the strong hygroscopic nature of glycerol, the water vapor present in the atmosphere is absorbed by the glycerol at the outer boundary of the rheometer at a mass vapor flux of $j_w$. The absorption of water by glycerol leads to a change in the fluid properties over time including a reduction in the fluid viscosity, which leads to a transient reduction in the effective viscosity measured by the system.\label{fig:setup}}
\end{figure*}

Previously, we reported that the strong hygroscopic nature of glycerol can cause unreliable viscosity measurements in a cone-and-plate rheometer due to the continuous vapor absorption from the outer edge \citep{shin2016benard}. Motivated by this observation, here we present a systematic study on the transient measurement of the viscosity of glycerol using a parallel-plate rheometer (see Figure \ref{fig:setup}).
As the secondary recirculating flow effectively disperses the absorbed water throughout the glycerol layer, we find that the rate of decrease of the measured viscosity is a complex function of the rheometer gap height, angular velocity, and the relative humidity. While the viscosity generally decreases over time as water is absorbed, we find that the rate of decrease is a non-monotonic function of the gap height. Using the theoretical solutions for the flow profile in a parallel-plate rheometer (i.e., Eq. (\ref{eq:savins})) along with numerical simulations, we show that this non-monotonic behavior is consistent with existing theory provided the gap height is not too small. In the limit of very small gap heights, we find that the behavior of the transient viscosity measurements is inconsistent with the existing theory for a parallel-plate rheometer. We hypothesize that misalignment effects in the rheometer result in non-negligible secondary flows at small gap heights that are responsible for this discrepancy in the measured viscosity data. By developing new theoretical solutions and computational simulations for the misaligned parallel-plate geometry, we show that this hypothesis is consistent with the measured viscosity data and that the misaligned rheometer model can predict the viscosity measurements across the full range of gap heights.

\section{Experimental viscosity measurements}

Here, we consider the transient viscosity measurements of glycerol in a parallel-plate rheometer where the outer fluid interface is exposed to the atmosphere. Due to the hygroscopic nature of glycerol, it will absorb water vapor from the atmosphere at the outer boundary as shown in Figure \ref{fig:setup}, which will subsequently lead to a local reduction in the viscosity of the fluid and a net reduction of the torque measured by the rheometer. Thus, the measured viscosity of glycerol in a parallel-plate rheometer is expected to decrease with time when exposed to atmosphere. Intuitively, the rate of decrease should depend on the water concentration/flux experienced at the outer boundary, which is influenced by the relative humidity in the atmosphere. The rate of decrease should also depend on the secondary recirculating flows in the rheometer, since these redistribute the relatively less viscous fluid where the water concentration is higher, thereby altering the stress distribution and total torque experienced by the upper plate of the rheometer.

\subsection{Experimental methods}
Glycerol was purchased from Sigma-Aldrich. Viscosity measurements were performed using a stress-controlled rheometer (Physica MCR 301, Anton Paar) with a parallel-plate configuration (plate diameter = 50 mm). The rheometer was placed inside an acrylic chamber in which the relative humidity (RH) was controlled using multiple vapor sources and a stream of dehumidified air. RH was constantly monitored using a digital hygrometer (VWR). Experiments were performed at 23$^\circ$C. 

\subsection{Experimental results}

\begin{figure*}
\centering\includegraphics[width=\textwidth]{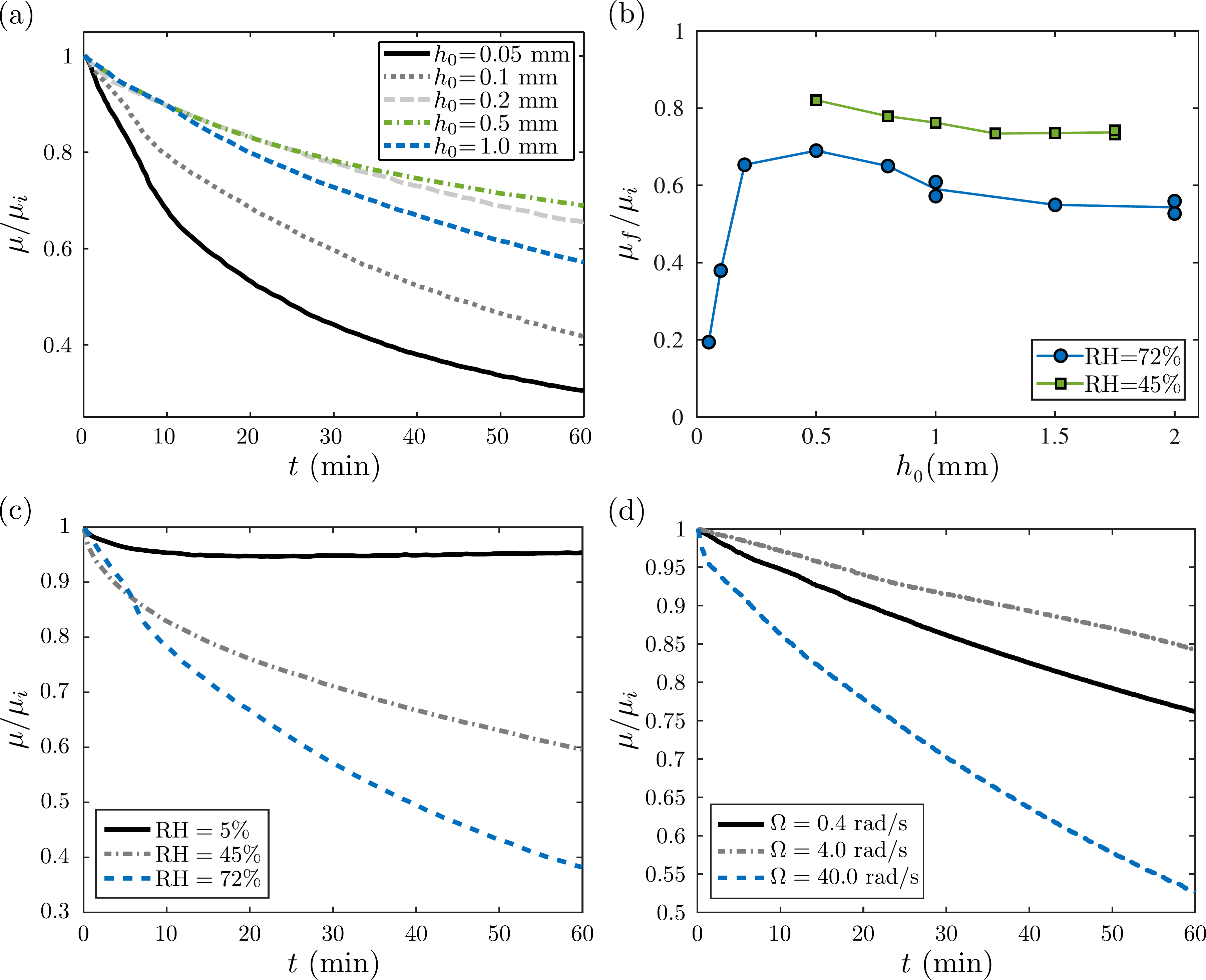}
\caption{Experimental viscosity measurements. (a) Transient decrease in the measured viscosity as a function of time, normalized by the initial viscosity $\mu_i$ with $\Omega=0.4$ rad/s and $R=2.5$ cm. Experiments were performed at a relative humidity of $\mathrm{RH}=53\%$. (b) Measured final viscosities $\mu_f$ of glycerol at $t=3600$ s normalized by $\mu_i$ with $\Omega=0.4$ rad/s and $R=2.5$ cm. Experiments were performed at relative humidities of $\mathrm{RH}=45\%$ and $\mathrm{RH}=72\%$. (c) Normalized transient viscosity measurements over time for varying relative humidities at a gap height of 0.1 mm and a rotation rate of 0.4 rad/s. (d) Normalized transient viscosity measurements over time for varying rotation rates at a gap height of 1 mm and a relative humidity of 54\%.}\label{fig:experimental}
\end{figure*}

The experimental results for the transient viscosity measurements are presented in Figure \ref{fig:experimental}. First, the transient viscosity measurements of glycerol in a relative humidity environment of 53\% are presented in Figure \ref{fig:experimental}a. The experiments were performed with a rheometer with plate radius $R=2.5$ cm at an angular velocity of $\Omega=0.4$ rad/s. Results are normalized by the initial viscosity $\mu_i$ to account for any small amount of water absorption by the glycerol while setting up the experiment. As can be seen, the measured viscosities decrease at varying rates depending on the gap thickness. These trends are all monotonic. However, the rate of viscosity decrease is seen to be a non-monotonic function of the gap thickness. That is, decreasing the gap height from $h_0=1.0$ mm to $h_0=0.5$ mm results in the viscosity dropping at a slower rate. However, subsequently decreasing the gap height results in the viscosity dropping at a faster and faster rate, until the viscosity drops sharply for the smallest gap height. Furthermore, experimental results for the final measured viscosity $\mu_f/\mu_i$ after $t=60$ minutes are shown in Figure \ref{fig:experimental}b as a function of gap height $h_0$ for two different relative humidities. Here, with $\text{RH}=72\%$ the viscosities are seen to drop faster than with $\text{RH}=45\%$ due to the increased mass flux of water vapor to the glycerol from the atmosphere. In addition, the final measured viscosity values also show the same non-monotonic behavior as in Figure \ref{fig:experimental}a. That is, decreasing $h_0$ first leads to an increase in $\mu_f/\mu_i$, followed by a sharp decrease for $h_0<0.5$ mm.

Next, the transient decrease in measured viscosity values are shown in Figure \ref{fig:experimental}c as functions of relative humidity at a rotation speed of 0.4 rad/s and a gap height of 0.1 mm. Here the results show a monotonic relationship in which increasing the relative humidity leads to a faster decrease in viscosity measurements due to the increased flux of water into the glycerol. Finally, the transient measured viscosities are shown in Figure \ref{fig:experimental}d for varying rotation speeds at a gap height of 1 mm and a relative humidity of 54\%. Here again a non-monotonic relationship with $\Omega$ is observed. At first, increasing the rotation speed from 0.4 rad/s to 4.0 rad/s leads to a slower decrease in viscosity, while subsequently increasing the rotation speed to 40.0 rad/s leads to a much faster decrease in the measured viscosity values. The key observations from the experimental measurements are:
\begin{enumerate}
\item Increased relative humidity leads to faster viscosity decrease for all measured cases.
\item The rate of viscosity decrease varies non-monotonically with the gap height. That is to say, for different gap heights, the measured viscosity decreases at a different rate over time.
\item There is a sharp decrease in the measured final viscosity for very small gap heights.
\end{enumerate}
Observation 1 above is a natural and intuitive result, since the hygroscopic nature of glycerol leads it to absorb more water from the atmosphere at higher humidities. Since the viscosity of a glycerol--water mixture varies monotonically with the water mass fraction, the rate of viscosity decrease can be expected to vary monotonically with the relative humidity. However, an intuitive explanation for observations 2 and 3 is not immediately obvious without a more careful consideration of the fluid dynamics and the coupled transport of dissolved water in the rheometer. In the following sections, we will seek a physical explanation for these two experimentally observed behaviors using theory and numerical simulations.

\section{Fluid dynamics of a glycerol--water mixture}

Before proceeding to analyze the specific fluid dynamics and transient viscosity evolution in a parallel-plate rheometer, in this section we first introduce all of the relevant governing physics that applies to such systems. This includes the governing Navier-Stokes equations for variable viscosity fluids, the advection-diffusion equation for the absorbed water concentration with variable diffusivity, the empirical relationships for the coefficients of viscosity and diffusivity of glycerol-water mixtures, as well as the saturation concentration of absorbed water in glycerol as a function of the relative humidity that will be used in the modeling.

\subsection{Navier-Stokes equations for a variable viscosity fluid}

In a glycerol-water mixture, the viscosity is strongly dependent on the local mass fraction of water. In systems with a homogeneous concentration of absorbed water, the constant viscosity form of the Navier-Stokes equations is appropriate. However, when gradients in water concentration exist in the system, due to circumstances such as mixing streams or the absorption of water at a gas-liquid interface, the viscosity of the fluid must be treated as a function of time and position. Note that for glycerol-water mixtures, the density is a weak function of the water concentration, ranging from approximately 1260 kg/m$^3$ for pure glycerol down to 1000 kg/m$^3$ for pure water. Here, we neglect this variation and use the density of pure glycerol, assuming the water mass fraction does not get too large. The corresponding continuity equation for an incompressible flow is simply $\nabla^*\cdot\boldsymbol u^* = 0$. In such a system, extra stresses arise in the fluid that are related to the gradients of viscosity, and the appropriate form of the Navier-Stokes equations (for a Newtonian constitutive equation) is:
\begin{equation}\label{eq:variableNS}
\nabla^*\cdot\boldsymbol u^* = 0     ~~~~\text{and}~~~~      \rho\frac{D\boldsymbol{u}^*}{Dt^*}=-\nabla^* p^* + \mu^* {\nabla^*}^2 \boldsymbol u^* + \nabla^*\mu^*\cdot\nabla^*\boldsymbol u^* + \nabla^*\mu^*\cdot\left(\nabla^*\boldsymbol u^*\right)^\mathsf{T},
\end{equation}
where $\rho$ is the fluid density, $\mu$ is the fluid viscosity, and $^\mathsf{T}$ denotes the transpose. Here, $^*$'s denote dimensional variable quantities. For the case of the flow in a rheometer system, we will represent the flow using cylindrical coordinates, and we nondimensionalize the governing equations with
\begin{equation}\label{eq:scalings}
r=\frac{r^*}{R},~~~z=\frac{z^*}{h_0},~~~u_r=\frac{u_r^*}{\Omega R},~~~u_\theta=\frac{u_\theta^*}{\Omega R},~~~u_z=\frac{u_z^*}{\Omega h_0},~~~p=\frac{p^*}{\mu_0 \Omega R^2/h_0^2},~~~\mu=\frac{\mu^*}{\mu_0},~~~\text{and}~~~t=\Omega t^*,
\end{equation}
where $\mu_0$ is a reference viscosity value. With these nondimensionalizations, the component form of Eq. (\ref{eq:variableNS}) in cylindrical coordinates is given by
\begin{subequations}\label{eq:variableNS_cylindrical}
\allowdisplaybreaks
\begin{align}
%
%
\nonumber&\Rey\left(\frac{\partial u_r}{\partial t}+u_r\frac{\partial u_r}{\partial r}+\frac{u_\theta}{r}\frac{\partial u_r}{\partial\theta}-\frac{{u_\theta}^2}{r}+u_z\frac{\partial u_r}{\partial z}\right)=-\frac{\partial p}{\partial r} +\mu\left(\frac{\epsilon^2}{r}\frac{\partial}{\partial r}\left(r\frac{\partial u_r}{\partial r}\right)+\frac{\epsilon^2}{{r}^2}\frac{\partial^2u_r}{\partial\theta^2}\right.\\
\nonumber&\qquad\left.+\frac{\partial^2u_r}{\partial {z}^2}-\epsilon^2\frac{u_r}{{r}^2}-\frac{2\epsilon^2}{{r}^2}\frac{\partial u_\theta}{\partial\theta}\right)+2\epsilon^2\frac{\partial\mu}{\partial r}\frac{\partial u_r}{\partial r} + \frac{\epsilon^2}{r}\frac{\partial\mu}{\partial\theta}\left(\frac{1}{r}\frac{\partial u_r}{\partial\theta}+\frac{\partial u_\theta}{\partial r}-\frac{u_\theta}{r}\right)\\
&\qquad + \frac{\partial \mu}{\partial z}\left(\frac{\partial u_r}{\partial z}+\epsilon^2\frac{\partial u_z}{\partial r}\right),\\
%
%
\nonumber&\Rey\left(\frac{\partial u_\theta}{\partial t}+u_r\frac{\partial u_\theta}{\partial r}+\frac{u_\theta}{r}\frac{\partial u_\theta}{\partial\theta}+\frac{u_\theta u_r}{r}+u_z\frac{\partial u_\theta}{\partial z}\right)=-\frac{1}{r}\frac{\partial p}{\partial\theta}+\mu\left(\frac{\epsilon^2}{r}\frac{\partial}{\partial r}\left(r\frac{\partial u_\theta}{\partial r}\right)+\frac{\epsilon^2}{{r}^2}\frac{\partial^2 u_\theta}{\partial\theta^2}\right.\\
\nonumber&\qquad\left.+\frac{\partial^2u_\theta}{\partial {z}^2}-\epsilon^2\frac{u_\theta}{{r}^2}+\frac{2\epsilon^2}{{r}^2}\frac{\partial u_r}{\partial \theta}\right)+\epsilon^2\frac{\partial\mu}{\partial r}\left(\frac{\partial u_\theta}{\partial r}+\frac{1}{r}\frac{\partial u_r}{\partial \theta}-\frac{u_\theta}{r}\right) + \frac{2\epsilon^2}{r}\frac{\partial\mu}{\partial\theta}\left(\frac{1}{r}\frac{\partial u_\theta}{\partial\theta}+\frac{u_r}{r}\right)\\
&\qquad + \frac{\partial \mu}{\partial z}\left(\frac{\partial u_\theta}{\partial z}+\frac{\epsilon^2}{r}\frac{\partial u_z}{\partial\theta}\right),\\
%
%
\nonumber&\Rey\,\epsilon^2\left(\frac{\partial u_z}{\partial t}+u_r\frac{\partial u_z}{\partial r}+\frac{u_\theta}{r}\frac{\partial u_z}{\partial\theta}+u_z\frac{\partial u_z}{\partial z}\right)=-\frac{\partial p}{\partial z}+\mu\left(\frac{\epsilon^4}{r}\frac{\partial}{\partial r}\left(r\frac{\partial u_z}{\partial r}\right)+\frac{\epsilon^4}{{r}^2}\frac{\partial^2u_z}{\partial\theta^2}+\epsilon^2\frac{\partial^2u_z}{\partial {z}^2}\right)\\
&\qquad + \frac{\partial\mu}{\partial r}\left(\epsilon^4\frac{\partial u_z}{\partial r}+\epsilon^2\frac{\partial u_r}{\partial z}\right) +\frac{1}{r}\frac{\partial\mu}{\partial\theta}\left(\frac{\epsilon^4}{r}\frac{\partial u_z}{\partial\theta}+\epsilon^2\frac{\partial u_\theta}{\partial z}\right) + 2\epsilon^2\frac{\partial\mu}{\partial z}\frac{\partial u_z}{\partial z},
\end{align}
\end{subequations}
where the Reynolds number is defined as $\Rey=\rho\Omega h_0^2/\mu_0$, and the gap aspect ratio is $\epsilon=h_0/R$, which is typically small in a parallel-plate rheometer. The corresponding continuity equation is given by
\begin{equation}\label{eq:continuity}
\frac{1}{r}\frac{\partial}{\partial r}\left(r u_r\right) + \frac{1}{r}\frac{\partial u_\theta}{\partial \theta} + \frac{\partial u_z}{\partial z} = 0.
\end{equation}
Along with boundary conditions, these equations govern the motion of variable viscosity fluids in a parallel-plate rheometer. The typical boundary conditions for such a system include no-slip conditions at the lower (stationary) and upper (rotating) plates, as well as a stress-free condition at $r=1$ at the gas-liquid interface.
\subsection{Water absorption and transport}

Along with the equations governing the fluid dynamics in the previous section, the system further requires a transport equation to model the absorption and transport of water in the glycerol. In particular, this transport can be modeled with an advection-diffusion equation that is given by
\begin{equation}\label{eq:advectiondiffusion_dim}
\frac{\partial c}{\partial t^*}=\nabla^* \cdot\left(D^*\nabla^* c\right) - \boldsymbol{u}^*\cdot {\nabla^*} c,
\end{equation}
where $c$ is the mass fraction of water in the glycerol, $D^*$ is the diffusivity of water in glycerol, and $\boldsymbol{u}^*$ is the dimensional fluid velocity vector. Here, $D^*$ is a spatially/temporally varying function of the local water concentration $c$. Using the same nondimensionalizations as above, Eq. (\ref{eq:advectiondiffusion_dim}) in cylindrical coordinates becomes
\begin{equation}\label{eq:solute_nondim}
\Pe\,\epsilon^2\frac{\partial c}{\partial t}=\frac{\epsilon^2}{r}\frac{\partial}{\partial r}\left(rD\frac{\partial c}{\partial r}\right)+\frac{\epsilon^2}{r^2}\frac{\partial}{\partial\theta}\left(D\frac{\partial c}{\partial\theta}\right)+\frac{\partial}{\partial z}\left(D\frac{\partial c}{\partial z}\right)-\Pe\,\epsilon^2\left(u_r\frac{\partial c}{\partial r}+\frac{u_\theta}{r}\frac{\partial c}{\partial \theta} + u_z\frac{\partial c}{\partial z}\right),
\end{equation}
where $D=D^*/D_0$ is the nondimensional diffusivity with reference value $D_0$ and $\Pe=\Omega R^2/D_0$ is the Peclet number representing the ratio of the timescale for diffusion of water in the radial direction to the convective timescale. Examining Eq. (\ref{eq:solute_nondim}), the diffusion in the $z$-direction is $\mathcal{O}(\epsilon^{-2})$ larger than the radial diffusion due to the separation in length scales: in the low-inertia, thin-gap limit, the water concentration will be approximately uniform in the $z$-direction.

Along with Eq. (\ref{eq:solute_nondim}), the absorbed water concentration must satisfy certain boundary conditions in the system. In particular, the concentration satisfies a no-flux condition at both the upper and lower plates of the rheometer (i.e. $\frac{\partial c}{\partial z}=0$ in the parallel-plate case). In addition, a boundary condition for the water concentration is needed at the outer glycerol/air interface. In general, this condition could be represented by a water flux condition such as by $-D\frac{\partial c}{\partial r}\bigr|_{r=1}=j_w$, where the flux of water $j_w$ could be a function of the local water concentration in the glycerol at the interface as well as the water vapor concentration and distribution in the air near the interface. The solution of such a flux will typically require also solving the water vapor transport problem in the surrounding environment, since these transport processes are coupled at the interface. For example, the transport of water vapor in the surrounding atmosphere may be affected by the rotation speed of the rheometer, which can drive flow in the surrounding air that may alter the vapor transport at the interface. Here, we neglect these effects and assume that the water vapor transport in the atmosphere is fast relative to the water concentration transport in the glycerol. This assumption is valid due to the substantially higher diffusivity of water vapor in air than absorbed water in glycerol, provided the recirculation in the rheometer is not significantly fast. Thus, we assume that the water mass fraction at $r=1$ instantaneously reaches its saturation value based on the local relative humidity in the surrounding air.

Considering both Equations (\ref{eq:variableNS_cylindrical}) and (\ref{eq:solute_nondim}), we see how the dynamics of the problem are fully coupled. In the fluid problem, the Navier-Stokes equations are coupled to the water transport problem via the dependence of viscosity on water mass fraction. The water transport problem is further coupled to the Navier-Stokes equations through the dependence on the flow velocity. Finally, the transport is further complicated by the dependence of diffusivity on water mass fraction. Due to this two-way coupling and the empirical nature of both the $\mu(c)$ and $D(c)$ relationships, it is difficult to seek theoretical solutions to the coupled dynamics. Thus, here we primarily rely on numerical simulations to solve the coupled transport problem.

\subsection{Physical properties of glycerol--water mixtures}

\begin{figure}
\centering\includegraphics[width=\textwidth]{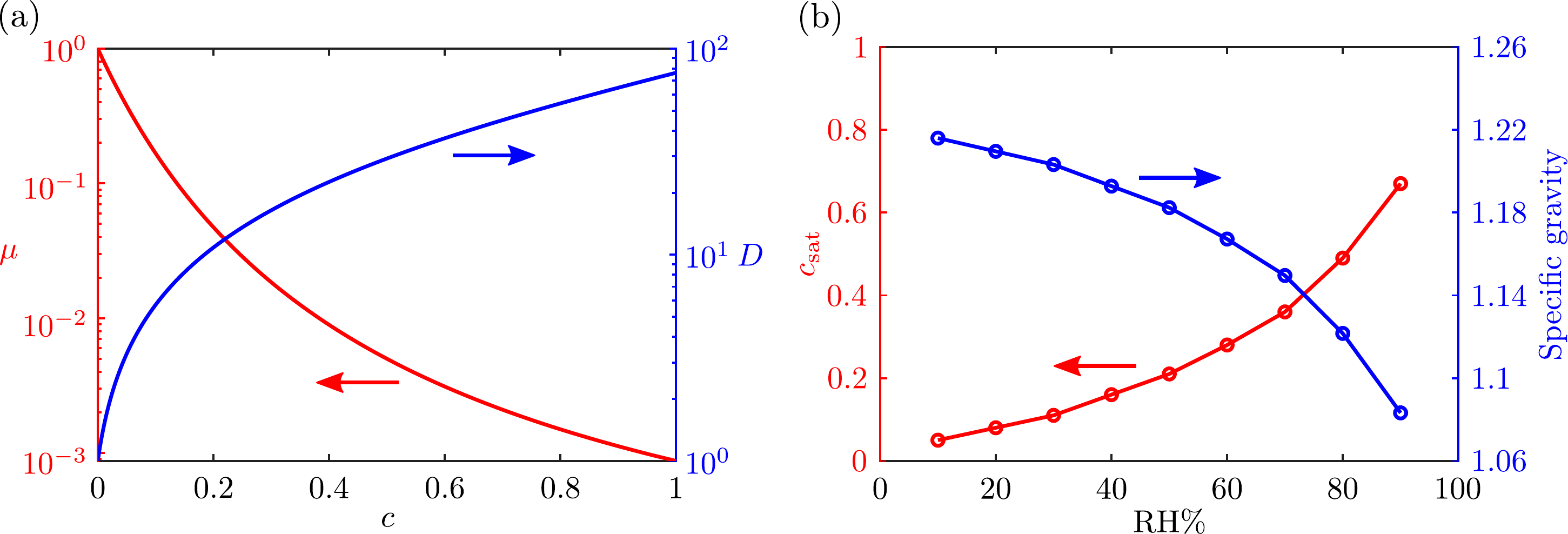}
\caption{(a) Empirical nondimensional viscosity and diffusivity relationships used in this study. (b) Saturation concentration of water in glycerol and corresponding specific gravity as functions of relative humidity.}\label{fig:physical_properties}
\end{figure}

Detailed empirical formulas for the viscosity of a glycerol--water mixture have been proposed by \citet{cheng2008formula} which are valid for water mass concentrations in the range of 0--100\% and for temperatures ranging from 0 to 100\degree C. The viscosity of a glycerol--water mixture at 22\degree C varies from around $\mu_g=1.1$ Pa$\cdot$s for pure glycerol down to $\mu_w=0.96$ mPa$\cdot$s for pure water, spanning a range of approximately three orders of magnitude. Here, we choose a reference viscosity value corresponding to that of pure glycerol $\mu_0=\mu_g$, such that the nondimensional viscosity varies from an initial condition of 1 down to as low as $\sim8.73\times10^{-4}$ if the water mass fraction were to approach 1. The empirical relationship determined between nondimensional viscosity and water concentration used here is shown in Figure \ref{fig:physical_properties}a.

The diffusivity of water in glycerol is also a function of the water mass fraction, and so $D^*$ is expected to evolve as a function of both position and time as more water is absorbed and transported throughout the system. An empirical relationship for the diffusivity of water in glycerol has also been developed for mixtures at 25 \degree C by \citet{d2004diffusion}, and is given by
\begin{equation}
D^*=\frac{1.024-0.91x_g}{1+7.5x_g}\cdot10^{-9}\,\frac{\text{m}^2}{\text{s}},
\end{equation}
where $x_g$ is the mole fraction of glycerol that is related to $c$ via
\begin{equation}
x_g=\frac{M_w(1-c)}{M_w(1-c)+M_gc},
\end{equation}
where $M_w$ is the molar mass of water and $M_g$ is the molar mass of glycerol. For reference, the diffusivity of water in pure glycerol ($c=0$) is $1.341\times10^{-11}$ m$^2$/s, which increases up to approximately $1.024\times10^{-9}$ m$^2$/s as the water mass fraction approaches 1. Here we again choose a reference value equal to the diffusivity in pure glycerol $D_0=D_g$, such that $D$ varies from 1 at $c=0$ up to approximately 76.4. This empirical relationship from \citet{d2004diffusion} is also shown in Figure \ref{fig:physical_properties}a. Finally, the saturation concentration of water in glycerol as a function of relative humidity is needed for the absorption boundary condition at $r=1$. These values were measured by the Glycerine Producers' Association and are given in Table 15 of \citet{glycerine1963physical}. For reference, these values are plotted in Figure \ref{fig:physical_properties}b along with the specific gravity as functions of relative humidity. With the full governing equations and empirical relationships for the physical parameters described above, we can now move on to consider the coupled fluid dynamics and water concentration transport in a rheometer. In the following section we first review the classical result for the axisymmetric parallel-plate rheometer with constant viscosity before moving on to cases with variable viscosity.
\subsection{Note on the assumption of constant density}
Before moving on, we briefly comment on the assumption of constant density. As water is absorbed into the glycerol, the resulting density gradients introduce the possibility for buoyancy-driven flows. An estimate for the magnitude of such effects is given by considering that a vertical change in density $\Delta\rho$ implies a radial pressure gradient on the order of $\Delta\rho g h_0/R$. We can balance this radial pressure gradient with a radial viscous stress gradient $\mu u_b/h_0^2$, where $u_b$ is the characteristic magnitude of the buoyancy-driven flow. Thus, we have $u_b\sim\frac{\Delta\rho g h_0^3}{\mu R}$. For buoyancy effects to be negligible, we need $u_b$ to be small relative to the magnitude of the inertial secondary-velocity components which are $\mathcal{O}\left(\frac{\rho\Omega^2h_0^2 R}{\mu}\right)$, as shown in Equation (\ref{eq:savins}). Thus, we need
\begin{equation}\label{eq:buoyancy_estimate}
\frac{\Delta\rho g h_0^3}{\mu R} \ll \frac{\rho\Omega^2h_0^2 R}{\mu}~~~~\longrightarrow~~~~\frac{\Delta\rho}{\rho} \ll \frac{\Omega^2 R^2}{g h_0}.
\end{equation}
The left-hand side of this inequality has a maximum value of around 0.2 when the saturation water concentration approaches 100\%, and so it will typically take a smaller value depending on the relative humidity. The right-hand side of the inequality depends on the system parameters, but a typical value with $\Omega = 4.0$ rad/s, $R=2.5$ cm, and $h_0=0.5$ mm is approximately 2.0. Thus, the buoyancy-driven flow can be expected to be less than 10\% of the inertial secondary flow. At larger gap heights and smaller rotation speeds, this inequality suggests that buoyancy effects become relatively more important, and could even dominate the dynamics. However, in the very slow rotation case, the dynamics are nearly 1D, such that the density variation is almost entirely in the radial direction and the negligible vertical density gradient should not affect the radial pressure gradient. So, Equation (\ref{eq:buoyancy_estimate}) should be considered only as an estimate of the importance of buoyancy effects.

\section{Classical result: Axisymmetric flow with constant viscosity}

Before moving on to consider the full coupled dynamics of glycerol absorbing water in a parallel-plate rheometer, we first review the classical Newtonian, constant viscosity flow solution in a parallel-plate rheometer. Understanding this flow is important because it illustrates the types of secondary flows we should expect in the rheometer, and also provides some initial insights into the non-monotonic relationship between measured viscosities and rotation speed and gap height in the full system. The axisymmetric parallel-plate rheometer has a well-known solution that has been previously described by multiple authors (see, for example \citet{middleman1968flow, bird1987dynamics}). Here, we briefly reproduce this calculation in our notation for consistency with later sections. We consider a parallel-plate rheometer with radius $R$ and gap thickness $h_0$. The lower plate is stationary and the upper plate rotates at a rotation speed of $\Omega$. We model the system using cylindrical coordinates with the origin located at the center of the bottom plate. Thus, the governing equations are the axisymmetric and constant viscosity forms of Eqs. (\ref{eq:variableNS_cylindrical}) and (\ref{eq:continuity}). The corresponding boundary conditions are no-slip at both the upper and lower plates, i.e. $(u_{r,\text{axi}},u_{\theta,\text{axi}},u_{z,\text{axi}})=(0,0,0)$ at $z=0$ and $(u_{r,\text{axi}},u_{\theta,\text{axi}},u_{z,\text{axi}})=(0,r,0)$ at $z=1$. For small gap heights $\epsilon\ll1$ a solution for the velocity and pressure can be sought in the form of an expansion in powers of $\epsilon^2$. Up to $\mathcal{O}(\epsilon^4)$, this axisymmetric solution is given by
\begin{subequations}\label{eq:axisymmetric}
\begin{align}
&u_{r,\text{axi}}(r,z)=-\frac{1}{12}r\Rey\,z(z-1)\left(-\frac{4}{5}+z+z^2\right)+\mathcal{O}(\epsilon^4),\\
&u_{\theta,\text{axi}}(r,z)=rz - \frac{r\Rey^2z}{6300}(8+z^3(35-63z+20z^3)) + \mathcal{O}(\epsilon^4),\\
&u_{z,\text{axi}}(z)=\frac{1}{30}\Rey\,z^2(z-1)^2(2+z)+\mathcal{O}(\epsilon^4),\\
&p_\text{axi}(r)=\frac{3r^2}{20}\Rey + \frac{1}{30}\Rey\,\epsilon^2 z(4-9z+5z^3) + \mathcal{O}(\epsilon^4).
\end{align}
\end{subequations}

Here, the subscript `axi' denotes the axisymmetric case, and the expression for $u_{r,\text{axi}}$ is equivalent to the result first presented by \citet{savins1970radial}, which was given above in dimensional form as equation (\ref{eq:savins}). The primary flow is the $u_{\theta,\text{axi}}=rz$ component with an $\mathcal{O}(\Rey^2)$ correction, while the leading-order secondary flows in the $r$- and $z$-directions are both $\mathcal{O}(\Rey)$. Since the flow of interest is axisymmetric, the primary velocity components of interest for redistributing absorbed species at the outer edge of the rheometer are the secondary velocity components, especially the radial component, since this will transport absorbed water from the outer edge inwards through the gap.

\begin{figure}
\centering\includegraphics[width=0.55\textwidth]{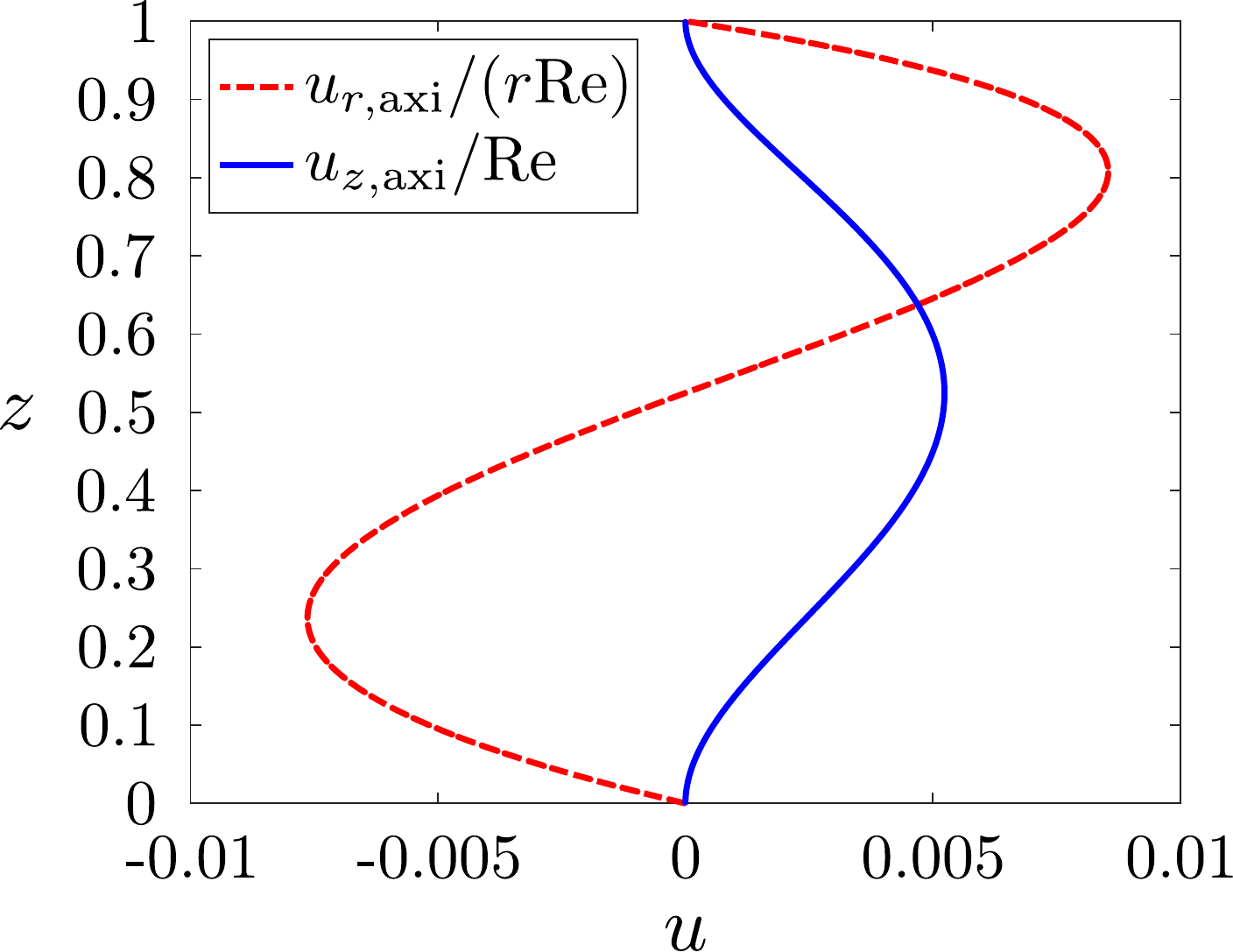}
\caption{Secondary velocity components $u_{r,\text{axi}}$ and $u_{z,\text{axi}}$ in a parallel-plate rheometer in the small-gap limit ($\epsilon\ll1$). The radial velocity proceeds outward along the upper-half of the gap, reverses at the outer edge, and proceeds radially inward along the lower-half of the gap. The $z$-component shows an upward drift along the middle of the gap that is independent of $r$. Both secondary velocities are $\mathcal{O}(\Rey)$. Results correspond to Eq. (\ref{eq:axisymmetric}).\label{fig:axisymmetric}}
\end{figure}

A visualization of the secondary velocity components is given in Figure \ref{fig:axisymmetric}. 
Here, the $u_{z,\text{axi}}$ component shows that there is an upward drift that is independent of $r$ throughout the rheometer. The radial component shows that in the upper-half of the gap the flow is directed radially outwards, and in the lower-half of the gap the flow is directed radially inwards. Keep in mind that the theoretical solution presented in Eq. (\ref{eq:axisymmetric}) must break down near the outer edge of the rheometer where $r\rightarrow1$, since the lubrication approximation fails in that region. In the true system, near the outer edge the outward-traveling flow in the upper-half of the rheometer must turn downwards for continuity and turn around to then travel radially inwards. The width of this turning region should be $\mathcal{O}(\epsilon)$, and thus is progressively more confined at the outer edge as the gap height decreases. This effect is not captured by Eq. (\ref{eq:axisymmetric}), although it may have an important role in the transport of absorbed water in the glycerol/water system. 

With this picture of the secondary flows, we can hypothesize an explanation for the non-monotonic behavior of the measured viscosity with gap height and rotation speed in the experimental results. First, in the slow rotation speed or small gap-height limit, absorbed water can only transport radially inwards via diffusion which is quite slow because the radial fluid velocity is $\mathcal{O}(\Rey)$. In reality, the diffusive transport will proceed faster than expected based on the diffusivity of water in pure glycerol, because the absorbed water increases the diffusion coefficient as it is absorbed. However, even using the diffusivity value that results for $c\rightarrow 1$, the radial diffusion remains a slow and inefficient process. Thus, in the low-$\Rey$ limit, the absorbed water remains highly confined near the outer edge of the rheometer, except over impractically long experimental timescales. On the one hand, this confines the relatively low viscosity fluid at the outer edge of the rheometer, where the torque is primarily generated, but it limits the amount of water that is absorbed since the concentration gradient is relatively diffuse at $r=1$. As $\Rey$ increases, the secondary flow begins to pull some of the absorbed water radially inward, steepening the gradient at $r=1$ and increasing the total flux of water into the system, while still leaving the absorbed water relatively confined at the outer edge. This results in a greater amount of absorbed water near the outer edge of the rheometer and a faster decrease in the measured viscosity. As $\Rey$ continues to increase, the secondary flows continue to more strongly pull the absorbed water away from the outer edge, further increasing the amount of absorbed water via the steeper gradient at $r=1$. However, it is plausible that above a certain $\Rey$ the secondary flows become sufficient enough to pull the absorbed water radially inwards a distance that is sufficiently far from the outer edge such that the net effect on the measured torque begins to lessen. That is, despite the fact that more water is absorbed, this low viscosity fluid is redistributed towards the inner part of the rheometer where the effect on the torque and measured viscosity is less. Thus, the constant viscosity, axisymmetric results can provide one possible explanation for the non-monotonic behavior seen between the measured viscosity  and the gap height and rotation speed. Furthermore, in the high-$\Rey$ limit, the absorbed water concentration can be expected to be well-mixed, promoting a rapid flux of water into the glycerol by maintaining a strong gradient at $r=1$ and rapidly redistributing the low viscosity fluid throughout the system. In such a regime, more care should be taken with determining the flux boundary condition, since the rate of water transport in the glycerol may approach the rate of water vapor transport in the outer flow problem where depletion of water vapor near the interface may limit the available flux into the glycerol. As a quick point of reference, with the definition of Reynolds number given by $\Rey=\rho_g\Omega h_0^2/\mu_g$, with the characteristic density and viscosity based on values for pure glycerol, the experimental results presented above in Figure \ref{fig:experimental} have Reynolds numbers ranging from $\sim1\times10^{-6}$ up to $\sim0.05$. While these values seem small, recall that the dimensional viscosity can vary by over three orders-of-magnitude, such that a locally defined Reynolds number could be significantly larger.

Finally, we note that the constant viscosity parallel-plate model is apparently inconsistent with the experimental observation of rapidly decreasing viscosity measurements at very small gap heights. In the limit of $\Rey\ll1$, the secondary flows in a parallel-plate rheometer are negligible. In this case, the transport equation simplifies to a purely 1D radial diffusion problem, and the evolution of water concentration becomes independent of both the gap height and rotation speed. Furthermore, the viscosity distribution likewise is independent of $h_0$ and $\Omega$, which is inconsistent with the sharp decrease in $\mu_f/\mu_i$ seen in Figure \ref{fig:experimental}b at very small gap heights. This will motivate us later in the paper to consider the potential role of misalignment. First, we examine in more detail the one-dimensional diffusive limit with variable viscosity.

\section{One-dimensional diffusive limit with variable viscosity}

First, we consider the evolution of the viscosity distribution and measured effective viscosity of glycerol absorbing water in the inertialess, one-dimensional diffusive limit. For small Reynolds numbers and gap heights, the axisymmetric form of Eq. (\ref{eq:solute_nondim}) becomes
\begin{equation}\label{eq:solute_nondim_1D}
\Pe\frac{\partial c}{\partial t}=\frac{1}{r}\frac{\partial}{\partial r}\left(rD\frac{\partial c}{\partial r}\right)+\frac{1}{\epsilon^2}\frac{\partial}{\partial z}\left(D\frac{\partial c}{\partial z}\right).
\end{equation}
Considering that the water concentration boundary condition at $r=1$ is independent of $z$, along with the no-flux conditions at the upper and lower plates, when $\epsilon\ll1$ it must be the case that $c$ is approximately independent of $z$, so that Eq. (\ref{eq:solute_nondim_1D}) further simplifies to
\begin{equation}\label{eq:solute_nondim_1D_reduced}
\Pe\frac{\partial c}{\partial t}=\frac{1}{r}\frac{\partial}{\partial r}\left(rD\frac{\partial c}{\partial r}\right),
\end{equation}
which is simply a 1D radial diffusion equation with variable diffusivity $D(c)$. In this regime, a better choice for the characteristic time scale would be the characteristic radial diffusion time $R^2/D_0$, the use of which would yield the same equation without the $\Pe$ factor on the left-hand side. For consistency, we continue to use the convective $1/\Omega$ timescale as the characteristic timescale. Here, the only boundary conditions that are needed are symmetry at $r=0$ and the saturation water mass fraction at $r=1$, i.e., $c(r=1)=c_\text{sat}$.

As the water concentration evolves, the anticipated viscosity measurement from the rheometer can be predicted through the use of the viscosity distribution as follows. A parallel-plate rheometer cannot measure the viscosity distribution throughout the fluid layer, but rather simply infers an effective viscosity $\mu_\text{eff}^*$ by measuring the total torque exerted on the upper plate as it spins. In dimensional form, the azimuthal velocity at small gap heights is $u_\theta^*=\Omega r^* z^*/h_0$. This velocity profile is valid regardless of the viscosity distribution since $c$ is a function of $r$ and $t$ only. The total torque experienced by the upper plate is then given by
\begin{equation}
T=\int\limits_0^{2\pi}\int\limits_0^R\frac{\mu^*\Omega}{h}{r^*}^3\, \mathrm{d}r^*\,\mathrm{d}\theta.
\end{equation}
If the viscosity is constant and uniform, the total torque on the upper plate is then
\begin{equation}\label{eq:torque}
T = \frac{\pi \mu^* \Omega R^4}{2h}~~~~\longrightarrow~~~~\mu^*=\frac{2hT}{\pi\Omega R^4}.
\end{equation}
The rheometer assumes a constant viscosity fluid and reports the ``effective'' viscosity of the fluid that is calculated from Eq. (\ref{eq:torque}) based on the measured torque. In the experimental system, the initial condition is assumed to be pure glycerol, such that $T_\text{init}=\pi\mu_g\Omega R^4/(2h)$, and we have
\begin{equation}
\frac{\mu_\text{eff}^*}{\mu_g}=\frac{T(t)}{T_\text{init}}=\frac{2h}{\pi\mu_g\Omega R^4}\int\limits_0^{2\pi}\int\limits_0^R\frac{\mu^*\Omega}{h}{r^*}^3\, \mathrm{d}r^*\,\mathrm{d}\theta ~~~~\longrightarrow~~~~ \mu_\text{eff}=\frac{2}{\pi}\int\limits_0^{2\pi}\int\limits_0^1 \mu{r}^3\, \mathrm{d}r\,\mathrm{d}\theta,
\end{equation}
which simplifies to
\begin{equation}
\mu_\text{eff} = 4\int\limits_0^1 \mu r^3\, \mathrm{d}r
\end{equation}
for axisymmetric flow.

\begin{figure}
\centering\includegraphics[width=\textwidth]{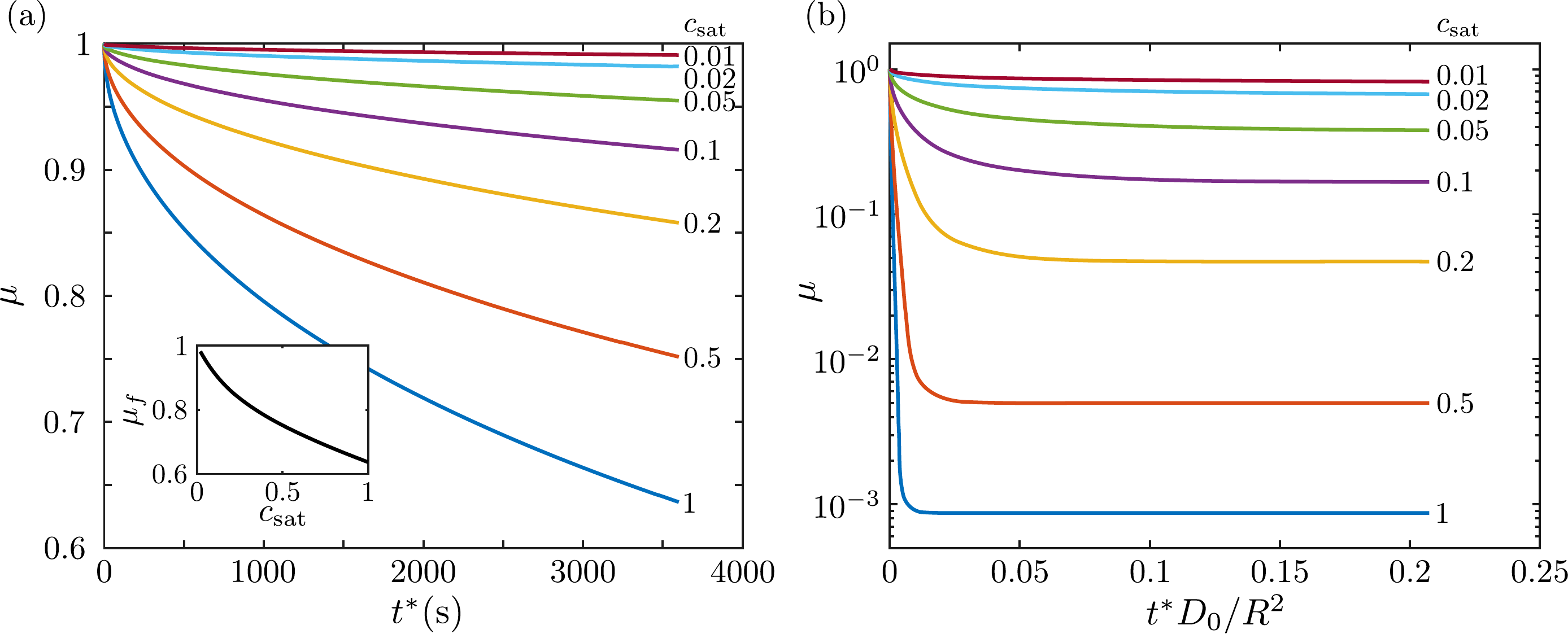}
\caption{Transient nondimensional viscosity measurements in the inertialess, 1D, axisymmetric regime. (a) Results simulated over 3600 seconds for comparison with experimental measurements. Results show substantial decreases in measured viscosities at large $c_\text{sat}$ values, but not as significant as those seen in the experiments. Figure inset shows the final nondimensional viscosity $\mu_f$ versus $c_\text{sat}$. (b) Results extended to much longer times to show the final saturation of the glycerol, which corresponds to the curves leveling off. Clearly, higher values of $c_\text{sat}$ reach saturation more quickly.}\label{fig:1D_results_unscaled}
\end{figure}

Here, we perform 1D transient simulations of Eq. (\ref{eq:solute_nondim_1D}) using the finite-difference method with second-order accuracy in space and first-order accuracy in time. Convergence studies were performed in space and time to verify the results. Using these simulations, we compute the effective nondimensional viscosity over time as water is absorbed at the outer edge and diffuses radially inwards. We perform these simulations over a range of $c_\text{sat}$ values which reproduces the effect of varying relative humidities. First, for comparison with experiments, the results are simulated for one hour to determine the degree of viscosity decrease that can be achieved via pure diffusion over the experimental timescale. These results are shown in Figure \ref{fig:1D_results_unscaled}a.
As can be seen, diffusion alone is sufficient to generate a significant decrease in measured viscosities over this timescale, although not to the degree seen in the experiments. For example, consider the experimental results in Figure \ref{fig:experimental}c, which were performed at a Reynolds number of $\Rey=\rho\Omega h_0^2/\mu_0=4.6\times10^{-6}$ and aspect ratio of $\epsilon=h_0/R=4\times10^{-3}$. Clearly, in such a regime the inertialess, 1D model would be expected to apply. However, the experimental results show a much larger decrease in viscosity over this timescale. The $RH=72\%$ results (corresponding to approximately $c_\text{sat}=0.386$) drop to around $\mu=0.38$, and the $RH=45\%$ results (corresponding to $c_\text{sat}=0.185$) drop to around $\mu=0.6$. However, in the 1D limit, the corresponding numerical predictions for $c_\text{sat}=0.4$ and $c_\text{sat}=0.2$ only decrease to around $\mu=0.78$ and $\mu=0.86$, respectively.

\begin{figure}
\centering\includegraphics[width=0.65\textwidth]{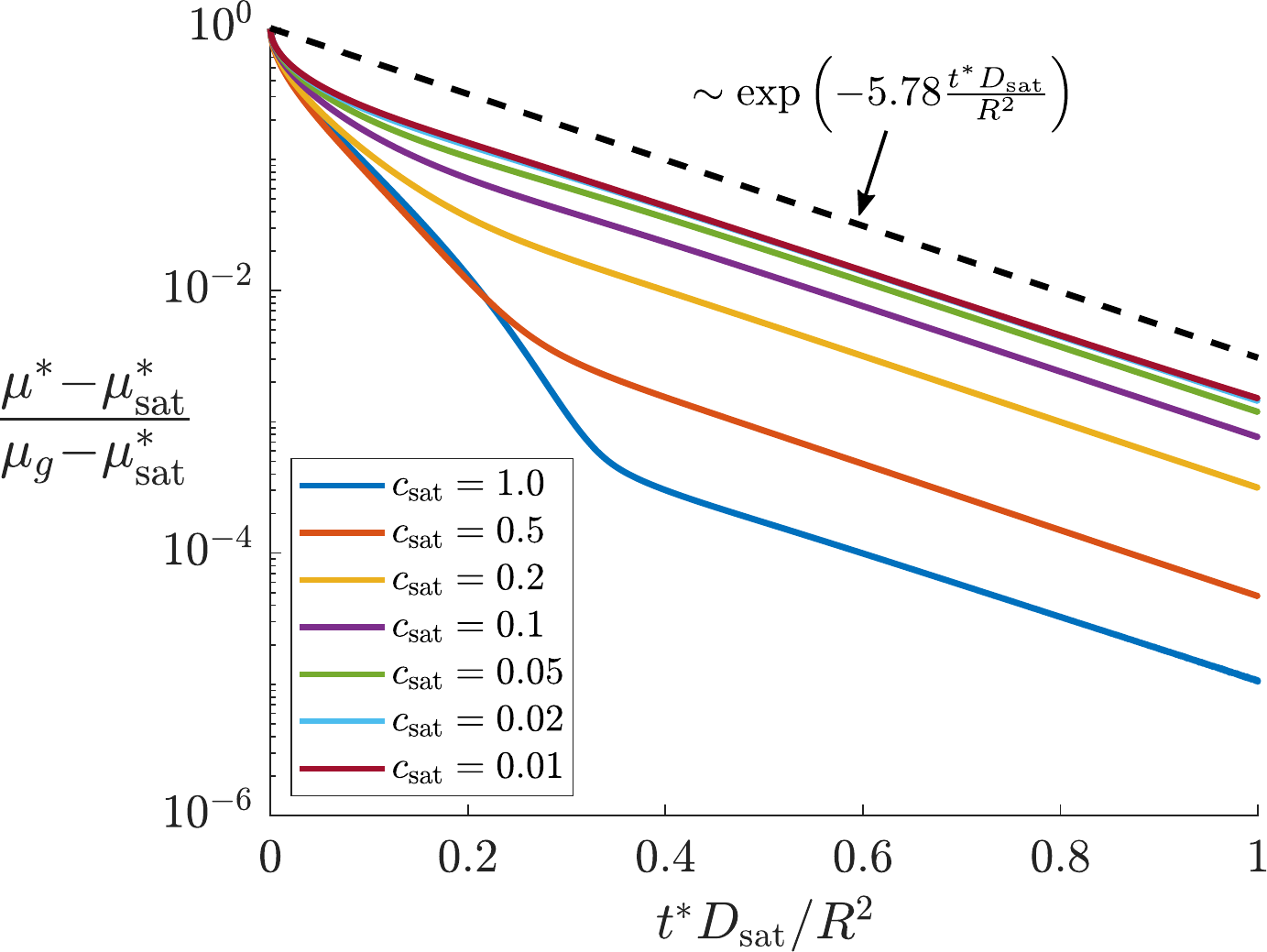}
\caption{Rescaled effective viscosities in the 1D, inertialess, axisymmetric limit. For each case the water concentration fully saturates approximately over the timescale $t^*=\mathcal{O}(R^2/D_\text{sat})$, which is consistent with diffusion primarily occurring at the saturation concentration diffusivity. At late times, the rescaled viscosities all approach the saturation values exponentially with a rate constant of 5.78, consistent with the 1D theory.}\label{fig:1D_results_rescaled}
\end{figure}

Thus, the experiments show a much larger decrease in viscosity over this timescale than the 1D model with pure diffusion. Furthermore, the results shown in Figure \ref{fig:1D_results_unscaled}a are clearly still evolving over this timescale, whereas in the long-time limit we expect all of the glycerol to homogenize at the saturation concentration based on the relative humidity. Therefore, we extend these results to much longer times in Figure \ref{fig:1D_results_unscaled}b, which shows the measured effective viscosities level off as the water concentration saturates. Here we see the influence of the variable diffusivity on the timescale for the diffusive process. With the characteristic diffusivity $D_0$, the timescale for the process would be expected to be $t^*=\mathcal{O}(R^2/D_0)$. However, as can be seen, most of the cases have fully saturated well before this timescale, especially at larger $c_\text{sat}$ values. This is due to the enhanced diffusion at larger water concentrations. In fact, a much better prediction for the timescale of this 1D diffusive process is to use the diffusivity based on $c_\text{sat}$, which we call $D_\text{sat}$. The rescaled results are shown in Figure \ref{fig:1D_results_rescaled},
which shows that for each case the water concentration in the glycerol has fully saturated over the timescale $t^*=\mathcal{O}(R^2/D_\text{sat})$. For each case, two regimes can be seen. In the early times, the water concentration in the glycerol is non-uniform, and so the diffusive transport in the domain proceeds with a spatially varying diffusivity coefficient. At late times, the water concentration throughout the system has nearly equilibrated at around the saturation concentration, such that the diffusion coefficient is nearly uniform and the results all decay exponentially with the same rate constant. This constant can be simply calculated by considering a 1D radial diffusion problem with constant diffusivity (since this is nearly the case at long times), where the transport in dimensional form is governed by
\begin{equation}
\frac{\partial c}{\partial t^*}=D_\text{sat}\frac{1}{r^*}\frac{\partial}{\partial r^*}\left(r^*\frac{\partial c}{\partial r^*}\right)~~~~\text{with}~~~~\frac{\partial c}{\partial r^*}\biggr|_{r^*=0}=0~~~~\text{and}~~~~c(r^*=R)=c_\text{sat}.
\end{equation}
The solution to this is given by
\begin{equation}
c(r^*,t^*)=c_\text{sat}+\sum\limits_{n=1}^\infty a_n \mathrm{e}^{-D_\text{sat}t^*\lambda_n^2} J_0(\lambda_n r),
\end{equation}
where the $a_n$ are coefficients that depend on the initial condition, and $J_0$ is the zeroth-order Bessel function of the first kind. The $\lambda_n$ eigenvalues here are the roots of $J_0$ divided by the radius $R$. Thus, at late times we see that $c-c_\text{sat}\sim \exp\left(-\chi_1^2 t^* D_\text{sat} / R^2\right)$, where $\chi_1=2.40483$ is the first root of the $J_0$ function. Thus we see the $\chi_1^2=5.7832$ exponential decay seen in Figure \ref{fig:1D_results_rescaled}.

The previous results are strictly valid in the inertialess ($\Rey\ll1$), small gap ($\epsilon\ll1$), and axisymmetric limits. These calculations are significantly simplified compared to the solution for the inertial regime since (1) the water concentration profile can no longer be assumed to be independent of $z$ due to the secondary velocity components and (2) the fluid velocity profiles must be recalculated continuously as the concentration profile evolves while taking into account the spatial variations in viscosity. Nevertheless, it is clear that we must extend our results to the inertial regime, since the 1D diffusion-dominated results cannot reproduce the same degree of viscosity decrease over the timescale of the experiments. These simulations are pursued in the following section.

\section{Inertial regime with variable viscosity}

Having explored the purely diffusion-dominated 1D axisymmetric regime in the previous section, we now extend our results to the inertial, axisymmetric regime. Recall that in the inertial regime, the coupled dynamics are governed by four dimensionless parameters, which are:
\begin{equation}
\Pe=\frac{\Omega R^2}{D_0},~~~\Rey=\frac{\rho\Omega h_0^2}{\mu_0},~~~\epsilon=\frac{h_0}{R},~~~\text{and}~~~c_\text{sat},
\end{equation}
whereas in the diffusion-dominated case the dynamics are governed only by $c_\text{sat}$. Thus, the system is governed by a relatively large parameter space. However, note that the ratio $\mu_0/(\rho D_0)=\mu_g/(\rho D_g)$ is fixed for glycerol, and the Peclet number can be written as
\begin{equation}
\Pe=\frac{\Omega R^2}{D_0}=\left(\frac{\rho\Omega h_0^2}{\mu_0}\right)\left(\frac{R^2}{h_0^2}\right)\left(\frac{\mu_0}{\rho D_0}\right)=\frac{\Rey}{\epsilon^2}\left(\frac{\mu_0}{\rho D_0}\right),
\end{equation}
so that the Peclet number is uniquely determined by the choice of $\Rey$ and $\epsilon$.
\subsection{Numerical methods}

\begin{figure}
\centering\includegraphics[width=\textwidth]{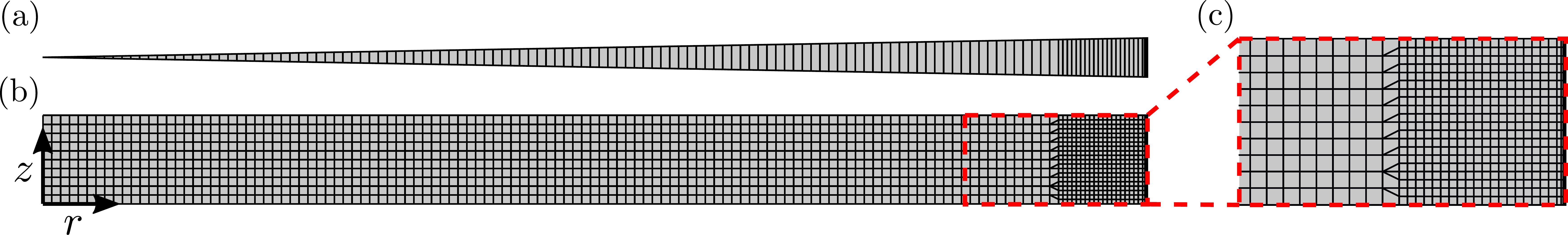}
\caption{Sample computational mesh design for the inertial, axisymmetric simulations. The grid has been coarsened by a factor of 3 in the $r$- and $z$-directions for visualization purposes. Local mesh refinement is used near $r=1$ to resolve the water concentration boundary layer. (a) Top-down view of the axisymmetric wedge mesh geometry. (b) Side view of the wedge mesh. (c) Zoom in of the local refinement near $r=1$. Several extra layers of very thin cells exist on the right-hand side which are difficult to see in order to resolve sharp concentration gradients that can occur at the boundary when inertial effects come into play.}\label{fig:mesh}
\end{figure}

Numerical simulations were performed using OpenFOAM \citep{Weller98} with an axisymmetric wedge-shaped mesh geometry with a wedge angle of 1\degree. Local mesh refinement was used near $r=1$ to resolve the water concentration boundary layer. A sample mesh design is shown in Figure \ref{fig:mesh}.
Simulations were performed using a custom in-house solver that iteratively updates the water concentration profile for 100 timesteps using a timestep of 0.01 seconds using second-order backward time-stepping and then recalculates the new steady-state velocity/pressure profiles using the SIMPLE algorithm \citep{barton1998comparison, jang1986comparison}. Thus, the solver assumes that the fluid velocity does not change much during one timestep. Convergence tests were performed to confirm that recalculating the velocity every 100 timesteps had a negligible impact on the calculated results compared to re-solving every timestep. The solver assumes that the velocity/pressure profiles are quasi-steady and only evolve when the water-concentration profile changes. The SIMPLE algorithm was used with relative pressure and velocity tolerances of $1\times10^{-5}$. Convergence tests also confirmed the results were insensitive to these tolerances. Finally, grid resolution convergence tests were performed, and a final base grid of $375\times 30$ cells in the $r\times z$ directions was chosen. The cells within the region from $r=1-2\epsilon$ to 1 were all further refined by one level. Finally, the final layer of cells at $r=1$ was further refined by halving three times. Using this grid, convergence tests indicate that the errors due to spatial discretization should be less than 1\%. Torque measurements were calculated by integrating the wall shear stress over the upper plate.

\subsection{Results}

\begin{figure}
\centering\includegraphics[width=\textwidth]{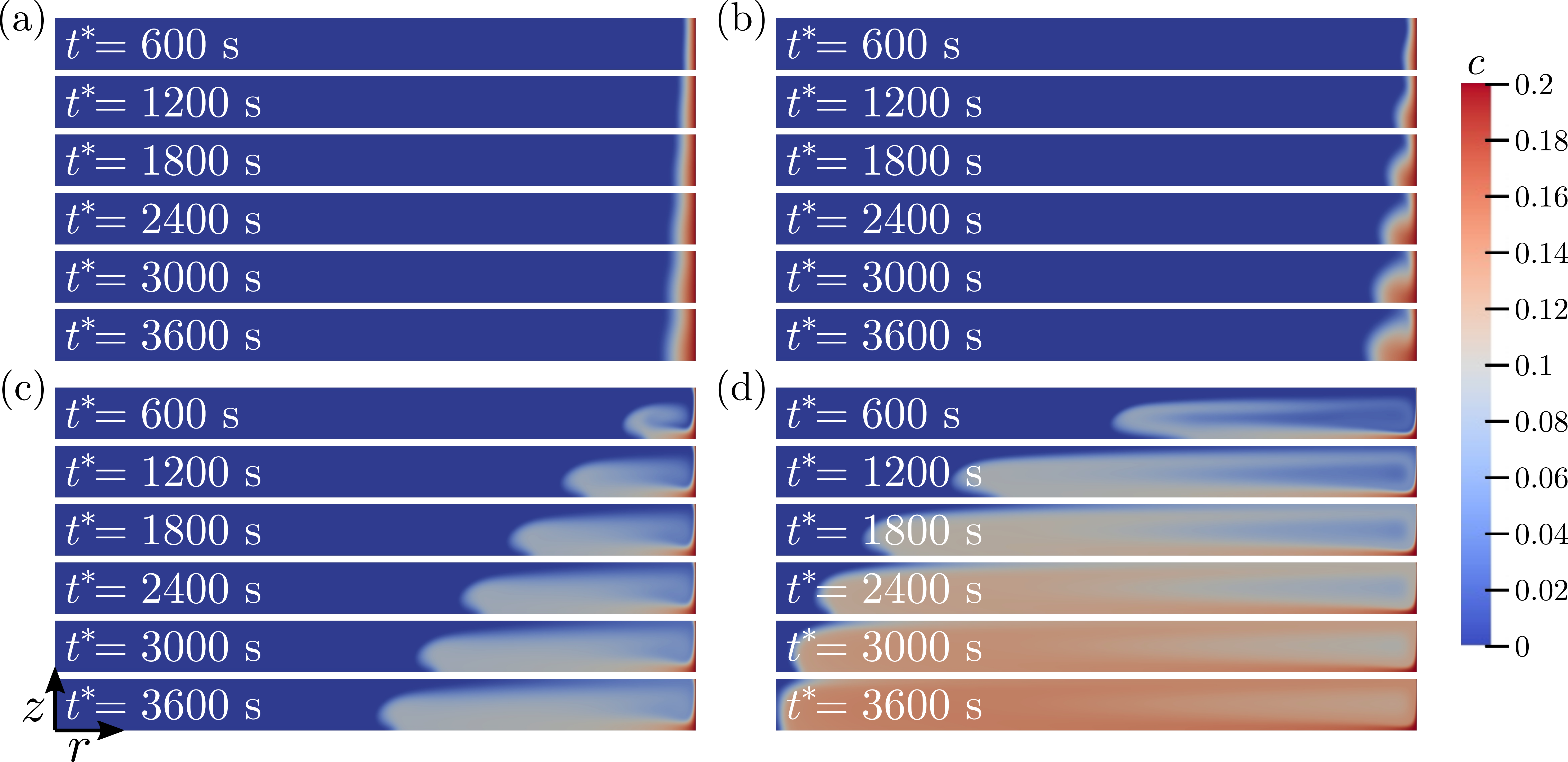}
\caption{Numerical results for the evolving water concentration profile $c$ over time for different rotation speeds $\Omega$ at $c_\text{sat}=0.2$, and $\epsilon=(1.0\times10^{-3}~\text{m})/(2.5\times10^{-2}~\text{m})=0.04$. Here, the angular speeds are (a) $\Omega=0.4$ rad/s, (b) $\Omega=1.0$ rad/s, (c) $\Omega=4.0$ rad/s, and (d) $\Omega=10.0$ rad/s. The corresponding nondimensional parameters are summarized in Table \ref{table:params}. Here, the Reynolds number (based on the saturation viscosity rather than $\mu_g$) ranges from $9.71\times10^{-3}$ up to 0.243 as the role of secondary (inertial) flows clearly grows with $\Omega$.}\label{fig:axisymmetric_varying_omega}
\end{figure}

Using the numerical methods described in the previous section, simulations were performed across a range of gap heights, $\Rey$, and relative humidities (through their proxy $c_\text{sat}$). Before introducing the final measured viscosity values for comparison with the experiments, we first present results illustrating the evolution and dynamics of the water concentration field in the glycerol over a range of gap heights and rotation speeds. A comparison of the evolving water concentration profiles at various rotation speeds is shown in Figure \ref{fig:axisymmetric_varying_omega}, and we give the results in dimensional form to make more clear the relationship of the changes to the experimental results presented earlier.
These simulations were performed at $c_\text{sat}=0.2$, and $\epsilon=(1.0\times10^{-3}~\text{m})/(2.5\times10^{-2}~\text{m})=0.04$ with rotation speeds of (a) 0.4 rad/s, (b) 1.0 rad/s, (c) 4.0 rad/s, and (d) 10.0 rad/s. The corresponding nondimensional parameters for these cases are summarized in Table \ref{table:params}.
\begin{table}
\centering\includegraphics[width=\textwidth]{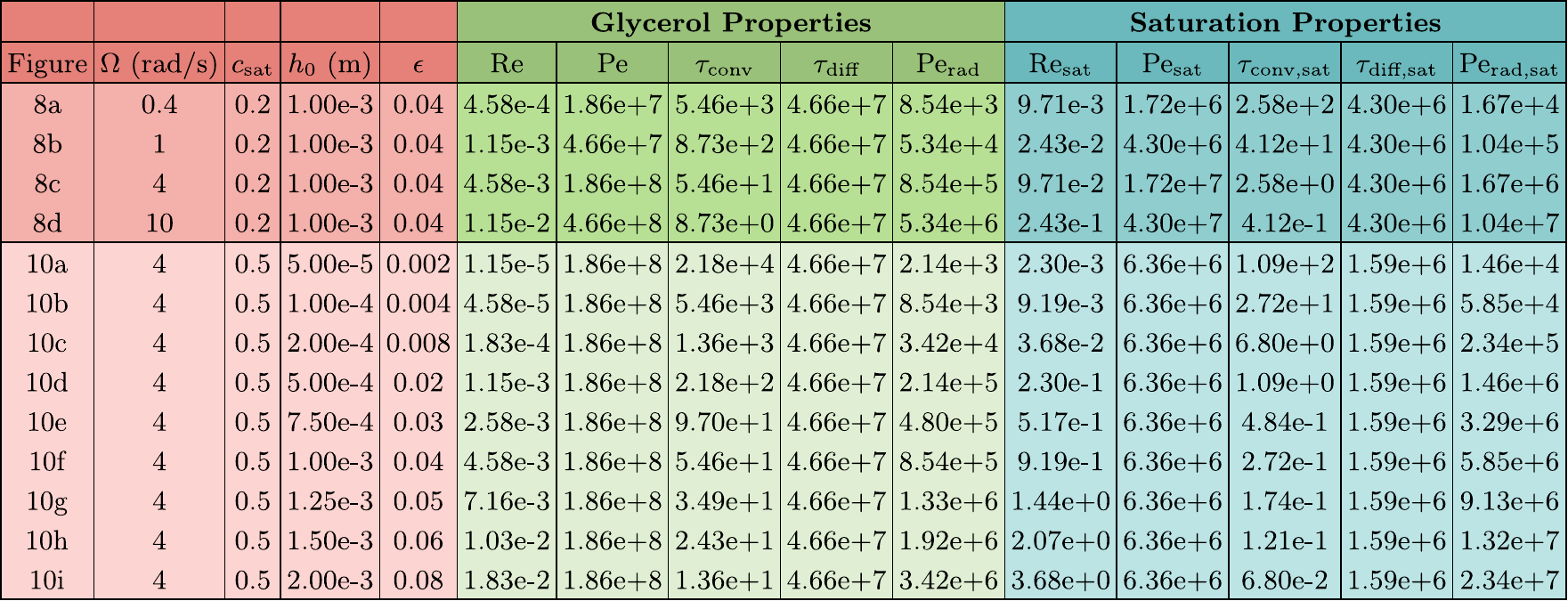}
\caption{Summary of the simulation parameters used in Figures \ref{fig:axisymmetric_varying_omega} and \ref{fig:axisymmetric_varying_h}. The timescale values $\tau_\text{conv}$ and $\tau_\text{diff}$ have units of seconds. Here, the parameters with `sat' subscripts are calculated based on the fluid properties at the appropriate saturation mass fraction of water, and parameters without this subscript are calculated based on the fluid properties of pure glycerol.}\label{table:params}
\end{table}
Here, the Reynolds number ranges from $\Rey=4.58\times10^{-4}$ up to $1.15\times10^{-2}$. This seems counter-intuitive, since even the smallest rotation speed case shows some deviation from a purely 1D, diffusive transport, as can be seen by the concentration variation in the $z$-direction, whereas the relatively small Reynolds numbers suggest inertial effects should be small for all of these cases. However, consider that $u_{r,\text{axi}}^*\sim\Omega R\Rey$. Then the characteristic time for convection in the radial direction is $\tau_\text{conv,rad}=(\Omega\,\Rey)^{-1}$, while the characteristic time for diffusion in the radial direction is $\tau_\text{diff,rad}=R^2/D_0$. Thus, an appropriate radial Peclet number is $\Pe_\text{rad}=\frac{\tau_\text{diff,rad}}{\tau_\text{conv,rad}}=\frac{\Omega R^2\Rey}{D_0}$. These values are also tabulated in Table \ref{table:params}. As can be seen, even for the smallest angular velocity case with $\Omega=0.4$ rad/s, the radial Peclet number is still greater than $\mathcal{O}(10^3)$, increasing up to $\mathcal{O}(10^6)$ at 10 rad/s. Thus, even at relatively small Reynolds numbers, the radial transport will be dominated by convection due to the relatively low diffusivity coefficients. For reference, Table \ref{table:params} also tabulates the nondimensional parameters based on the viscosity and diffusivity values associated with the saturation concentration $c_\text{sat}$ rather than reference values based on pure glycerol. Here, the Reynolds numbers are increased while both the convective and diffusive timescales are decreased due to the reduced viscosity and increased diffusivity at increased water concentrations.

\begin{figure}
\centering\includegraphics[width=\textwidth]{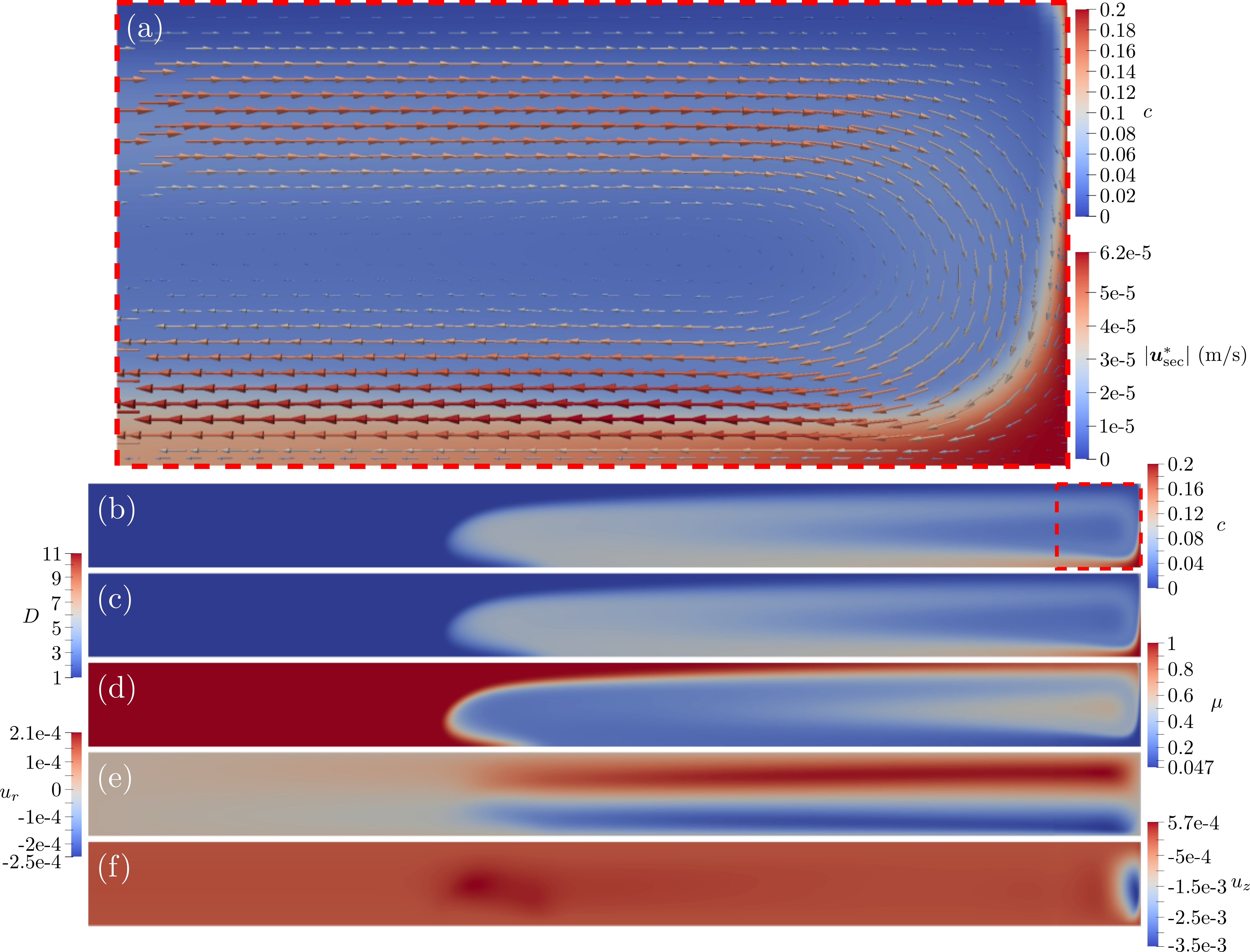}
\caption{Detailed look at the glycerol/water dynamics for the case corresponding to Figure \ref{fig:axisymmetric_varying_omega}d taken at $t=1\times10^4$ (see Table \ref{table:params} for all relevant parameters. (a) Secondary velocity vectors colored and scaled by the magnitude of the secondary velocity superimposed on a colormap of the water concentration profile. As can be seen, the water begins to diffuse inward from the outer boundary, where the secondary flow pulls the absorbed water downward and then radially inward along the bottom plate, leading to a steep concentration gradient at the outer edge as the rotation speed is increased. (b) Full water concentration profile over a full axisymmetric cross-section. (c-f) Nondimensional diffusivity coefficient, viscosity, radial velocity component, and $z$ velocity component.}\label{fig:axisymmetric_details}
\end{figure}

Examining the transport dynamics in Figure \ref{fig:axisymmetric_varying_omega}, we see that the transport of water concentration is consistent with the axisymmetric, constant viscosity flow picture described above. In particular, in the constant viscosity case, flow proceeds radially outward along the upper plate, turns downward, and then flows radially inward along the lower plate. This is shown in more detail in Figure \ref{fig:axisymmetric_details}a.
Here, the arrows are color-coded and scaled by the magnitude of the secondary velocity components $\left|\boldsymbol u^*_\text{sec}\right|$, and the background is color-coded by the water concentration profile. Here, $\boldsymbol u^*_\text{sec}$ is the velocity field on the slice in the $r$- and $z$-direction. As can be seen, the water begins to diffuse inwards from the outer edge, but then the secondary velocity pulls the absorbed water down along the outer edge and then radially inwards along the bottom plate. This creates a very thin boundary layer region near the outer edge of the rheometer, which gets thinner as $\Pe_\text{rad}$ increases. The full solute concentration profile corresponding to Figure \ref{fig:axisymmetric_details}a is shown in \ref{fig:axisymmetric_details}b, and the corresponding dimensionless diffusivity, viscosity, radial velocity, and $z$ velocity are shown in \ref{fig:axisymmetric_details}c-f, respectively. This case corresponds to the parameters previously shown in Figure \ref{fig:axisymmetric_varying_omega}d with the parameters shown in Table \ref{table:params}, and all results are at the nondimensional time $t=\Omega t^*=(10~\text{rad/s})(1000~\text{s})=1\times10^4$.

As can be seen in Figure \ref{fig:axisymmetric_details}, the regions of high water concentration correspond to the regions of increased diffusivity and decreased viscosity. The radial velocity component resembles the flow for the axisymmetric constant viscosity case with outward radial flow along the upper half of the domain and inward radial flow along the lower half, except that the magnitude of both is increased throughout the extent of the low viscosity region. Furthermore, at the front of the propagating front of water concentration, there is a steep gradient in viscosity that corresponds to an upward secondary velocity due to the viscosity gradients as seen in Figure \ref{fig:axisymmetric_details}d,f.

\begin{figure}
\centering\includegraphics[width=\textwidth]{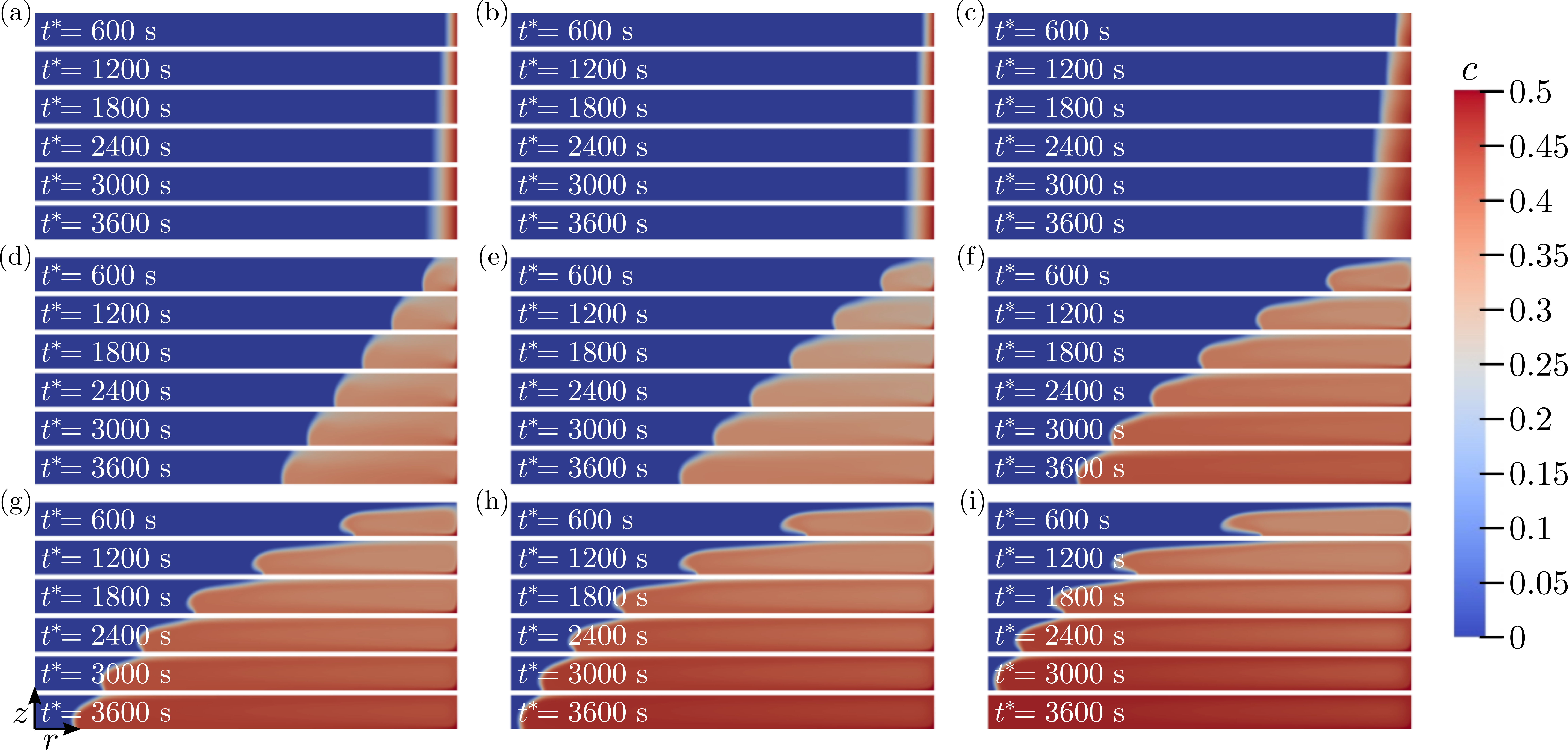}
\caption{Numerical results for the evolving water concentration profile $c$ over time for different gap heights $h_0$ at $c_\text{sat}=0.5$, and $\Omega=4.0$ rad/s. Here, the gap heights are (a) 0.05  mm, (b) 0.1 mm, (c) 0.2 mm, (d) 0.5 mm, (e) 0.75 mm, (f) 1.0 mm, (g) 1.25 mm, (h) 1.5 mm, and (i) 2.0 mm. The corresponding nondimensional numbers are summarized in Table~\ref{table:params}. Over this parameter range, the Reynolds numbers (based on the saturation viscosity) range from $2.30\times10^{-3}$ up to 3.68 and the gap aspect ratio ranges from 0.002 to 0.08. Thus, these cases capture the full transition from the 1D, diffusive limit, up to the inertial regime.}\label{fig:axisymmetric_varying_h}
\end{figure}

Finally, a last illustration of the water concentration dynamics at higher $c_\text{sat}$ is presented in Figure \ref{fig:axisymmetric_varying_h}
as a function of gap height. These results show the evolution of the water concentration profile over time at $c_\text{sat}=0.5$ and $\Omega=4.0$ rad/s for gap heights ranging from 0.05 mm to 2.0 mm. Again, all of the relevant nondimensional parameters are given in Table \ref{table:params} based on both the pure glycerol reference values and the saturation values. Here, the Reynolds number based on the saturation parameters ranges from $2.30\times10^{-3}$ up to 3.68, representing a transition from the inertialess regime into the moderate inertial regime. The radial Peclet numbers based on the saturation properties remain relatively high, increasing from $1.46\times10^4$ up to $2.34\times10^7$ as the gap height increases, suggesting that the radial transport of water is dominated by convection in these regimes. Nevertheless, the transport at the smallest gap heights is approximately 1D, suggesting that the $\Pe_\text{rad}$ threshold for this transition is at a relatively large magnitude in this particular system.

\begin{figure}
\centering\includegraphics[width=0.5\textwidth]{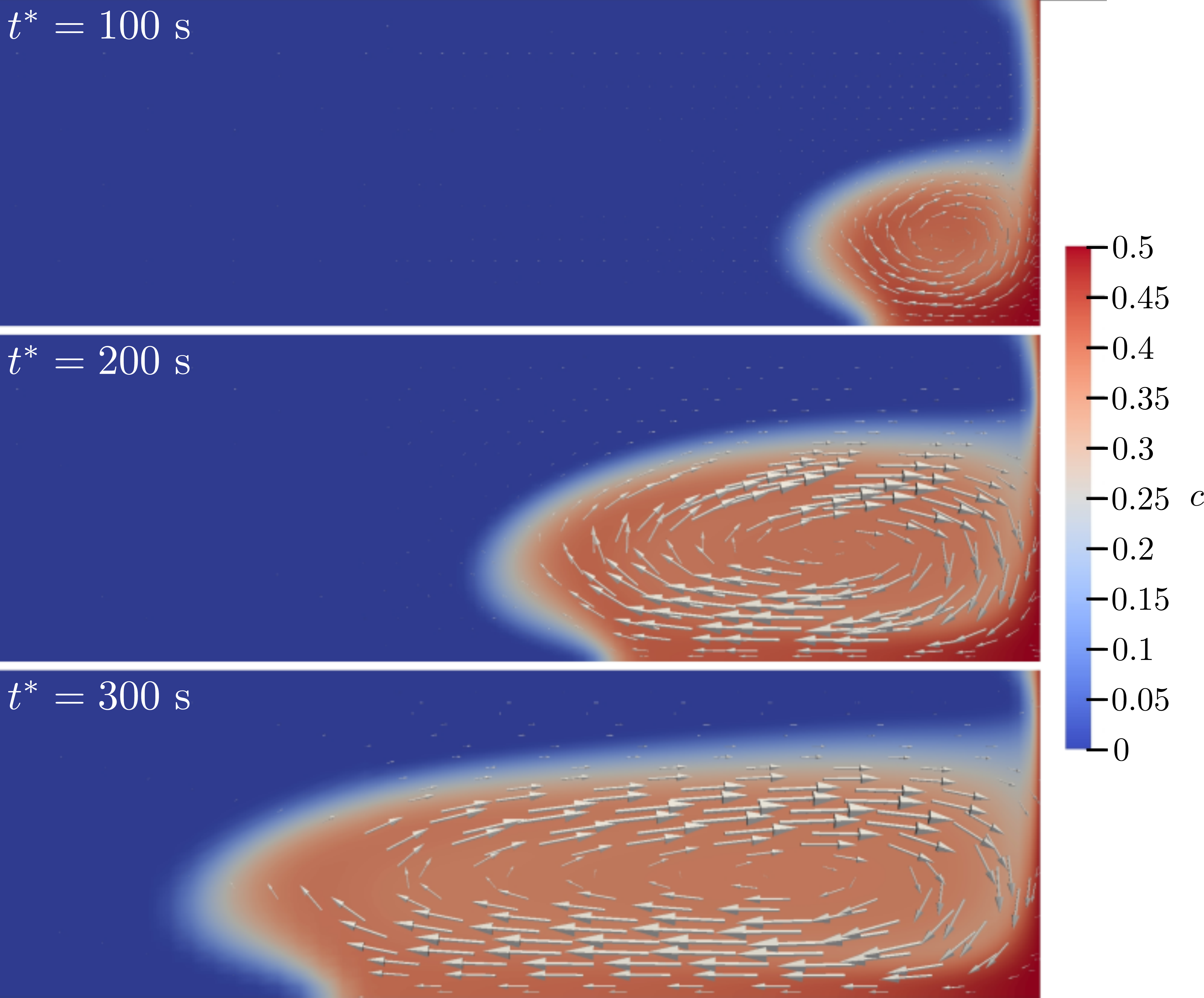}
\caption{Enhanced secondary recirculation in the low-viscosity region corresponding to larger $c_\text{sat}$ values. These results illustrate the enhanced mixing effect that is seen at early times with $c_\text{sat}=0.5$ for the parameters shown in Figure \ref{fig:axisymmetric_varying_h}f. At large values of $c_\text{sat}$, the local viscosity drops in regions of large $c$ to such a degree that the local recirculation dominates the expected secondary motions for constant viscosity, axisymmetric flow. }\label{fig:enhanced_mixing}
\end{figure}

Note that the qualitative picture of the water concentration evolution is different in Figures \ref{fig:axisymmetric_varying_omega} and \ref{fig:axisymmetric_varying_h}. In particular, Figure \ref{fig:axisymmetric_varying_h}f corresponds to the same gap height and rotation speed as Figure \ref{fig:axisymmetric_varying_omega}c, except with an increased $c_\text{sat}$ value of 0.5 versus 0.2, respectively. While this seems like a relatively minor change, the corresponding $\mu_\text{sat}$ value is an order of magnitude smaller at $c_\text{sat}=0.5$, which leads to an order of magnitude stronger secondary flows in the region of locally low viscosity. This generates an enhanced mixing that leads to a more homogeneously propagating front of water concentration. This can be visualized in Figure \ref{fig:enhanced_mixing}
for the case corresponding to Figure \ref{fig:axisymmetric_varying_h}f at early times. As can be seen, the local secondary recirculation in the low-viscosity region completely dominates the expected global secondary recirculation for the axisymmetric, constant viscosity case. In fact, that global velocity field (the axisymmetric constant viscosity solution) is negligible on the figure. This enhanced local recirculation at larger $c_\text{sat}$ values explains why the water propagates as a more uniform front in that regime, as opposed to being pulled down and inward along the lower plate as was illustrated in Figure \ref{fig:axisymmetric_details}a for a smaller $c_\text{sat}$ value.

\begin{figure}
\centering\includegraphics[width=\textwidth]{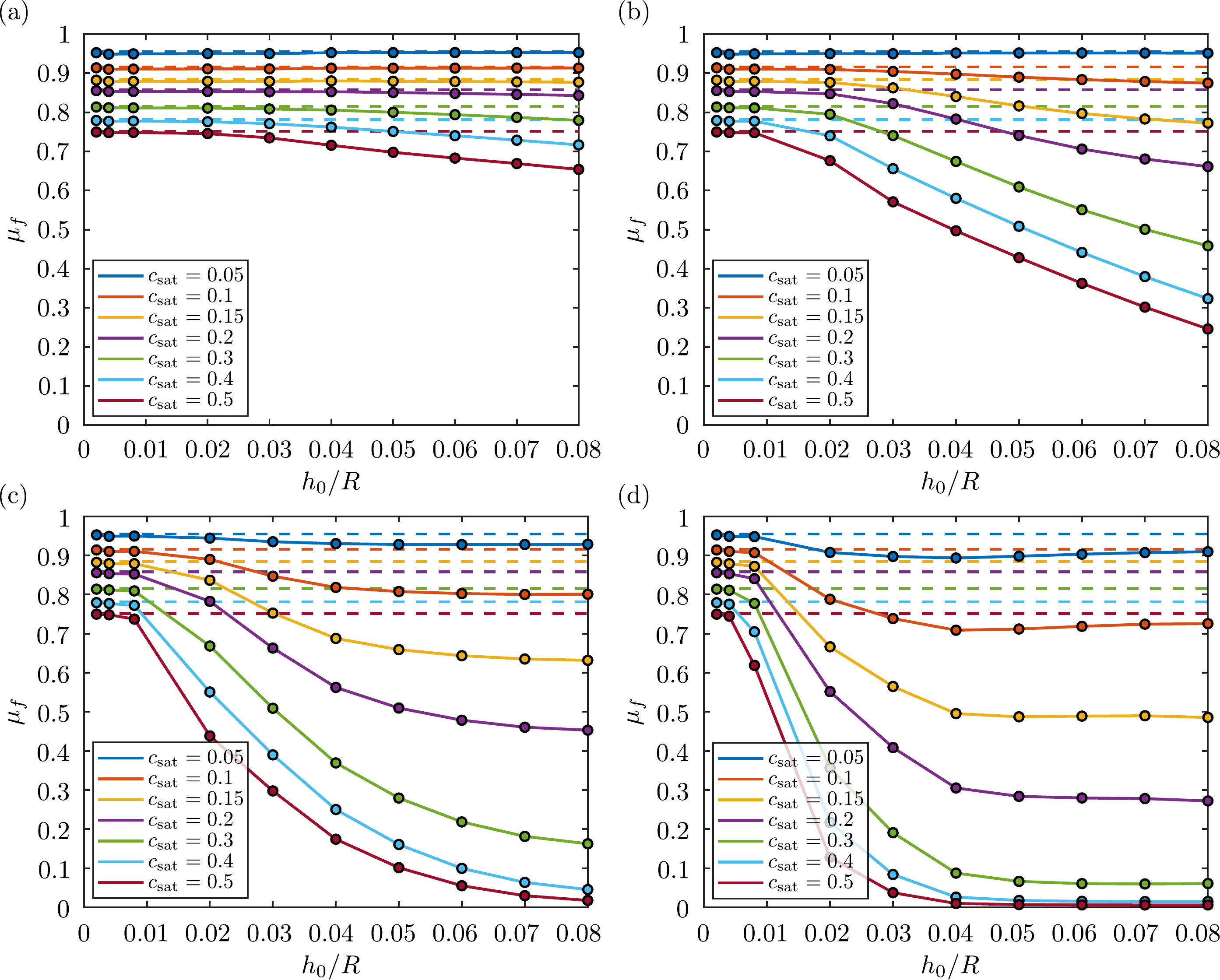}
\caption{Compilation of all measured final dimensionless viscosities $\mu_f$ at $t^*=3600$ s from the axisymmetric, inertial, variable viscosity simulations for comparison with the experimental results. Dashed lines indicate the 1D, inertialess, diffusion-dominated results described in the previous section. Results are plotted separately by rotation speed with values of (a) 0.4 rad/s, (b) 1.0 rad/s, (c) 2.0 rad/s, and (d) 4.0 rad/s. Clearly, deviations from the diffusion-dominated limit increase with gap height and angular rotation speed due to the increase of inertial secondary flows, as well as with increasing $c_\text{sat}$ due to local reductions in viscosity (and consequent increases in inertial effects).}\label{fig:axisymmetric_compiled}
\end{figure}

Finally, having characterized and visualized the coupled transport dynamics in the parallel-plate, axisymmetric, inertial regime, we now calculate the measured effective viscosities in these simulations to see if the proposed model can fully capture the trends seen in the experimental results. The final measured dimensionless viscosities $\mu_f$ at $t^*=3600$~s are presented in Figure \ref{fig:axisymmetric_compiled}
as functions of gap aspect ratio $\epsilon$, saturation concentration $c_\text{sat}$, and angular rotation speed $\Omega$. The corresponding angular rotation speeds in the figure are (a) 0.4 rad/s, (b) 1.0 rad/s, (c) 2.0 rad/s, and (d) 4.0 rad/s. In the figure, the dashed lines correspond to the predictions of the 1D, axisymmetric, diffusion-dominated results previously described in Figures \ref{fig:1D_results_unscaled} and \ref{fig:1D_results_rescaled}. Note that in the experiments, the reported viscosities were nondimensionalized by $\mu^*_i$, the initial measured viscosity at $t=0$, and in the simulations the reported viscosities have been nondimensionalized by $\mu_g$. Here, several key relationships and trends emerge from the results. First, we see clearly that in every case, the results approach the 1D diffusion-dominated limit as $\epsilon\rightarrow0$ for constant $\Omega$, and they appear to also approach this limit as $\Omega\rightarrow0$ for constant $\epsilon$. For a given $\Omega$, deviations from this limit increase as $\epsilon$ increases, due to enhanced inertial effects, as well as for increased $c_\text{sat}$. This latter trend is also due to an increase in inertial effects, although indirectly through a decreasing in the local viscosity. Furthermore, increasing $\Omega$ clearly leads to more significant deviations from the 1D limit due to increasing secondary inertial flows.

Comparing these results with the experimental results, the axisymmetric inertial simulations do seem to capture many features of the experimental results. In particular, we generally see decreased $\mu_f$ values at larger gap heights and larger $c_\text{sat}$ values (i.e., RH values), which are consistent with Figure \ref{fig:experimental}. Specifically, the numerical results in Figure \ref{fig:axisymmetric_compiled}a correspond to the same rotation speed ($\Omega=0.4$ rad/s) as Figure \ref{fig:experimental}b. Similar trends are seen (except at small gap heights, which will be discussed below), with slightly less significant decreases in viscosity in the simulations compared to the experiments. Increasing the rotation speed to 1.0 rad/s in the simulations shows more significant viscosity decreases than the experimental results at 0.4 rad/s. So the experimental results at 0.4 rad/s agree well quantitatively with numerical predictions slighty above 0.4 rad/s. One trend seen in the experiments that we do not see in the simulations is the non-monotonic relationship between viscosity decrease and angular rotation rate seen in Figure \ref{fig:experimental}d. However, we do see evidence of a non-monotonic relationship with increasing inertial effects in the simulations. In particular, at $\Omega=4.0$ rad/s with $c_\text{sat}=0.1$ (orange curve in Figure \ref{fig:axisymmetric_compiled}d), the final measured viscosity first decreases and then increases with increasing gap height, demonstrating that the axisymmetric case can demonstrate such trends.

The most significant experimental result that these simulations cannot explain is the large decrease in measured viscosity values at small gap heights. In fact, one of the most consistent results of the axisymmetric, inertial simulations is the approach to the 1D, diffusion-dominated regime at small gap heights for any angular rotation speed. Thus, these axisymmetric simulations apparently fail to account for some effect that becomes dominant at small gap heights. We hypothesize that this is due to misalignment effects that only become significant at very small gap heights in practical parallel-plate rheometers. In the next section, we perform additional simulations based on a misaligned geometry in an attempt to validate this hypothesis.

\section{Role of misalignment}

In the previous section, we clearly saw that the axisymmetric, inertial, variable viscosity model fails to account for the sharp decrease in measured viscosity at small gap heights. Thus, we must consider what possible sources of error could account for these effects. A variety of experimental challenges exist for performing accurate measurements with a rheometer, such as underfilling of the parallel-plate gap, instrument inertia, and surface tension effects \citep{hellstrom2014errors, ewoldt2015experimental}. In addition to these, there are practical sources of error associated with the mechanical uncertainties in the rheometer itself. A key source of these errors comes from deviations in the geometry of the gap containing the fluid. These errors in the gap geometry could arise from non-parallelism, non-concentricity, non-flatness of the plates, non-zero slip lengths at the upper or lower plates, edge effects at the outer edge of the rheometer, or errors in the gap-zeroing procedure \citep{connelly1985high, kramer1987measurement, kalika1989gap, henson1995effect, davies2005gap, andablo2010method, andablo2011nonparallelism}. One reason the discussion of these sources of error arose was due to the experimental observation that as gaps decreased below several hundred microns, measured viscosities began to have systematic errors, typically decreasing with the gap height as also shown in Figure \ref{fig:experimental}b \citep{walters1975rheometry, pipe2009microfluidic}.

Based on these observations, a variety of studies suggest that a key factor in this discrepancy in our measurements and simulations could be the misalignment of the rotating plate. Although it is commonly assumed that the plates are perfectly aligned, a number of reports indicate that a small but finite misalignment is prevalent in parallel-plate and cone-and-plate rheometers \citep{andablo2011nonparallelism}. In fact, the gap height can vary over 50~$\mu$m across a few centimeters in a parallel-plate rheometer due to the non-parallelism in the gap, causing a significant error in the viscosity measurements in narrow-gap, high-shear-rate experiments \citep{pipe2008high,davies2008thin}. This is due to the fact that the misalignment introduces additional lubrication forces in the fluid layer. A variety of semi-empirical techniques have been developed to account for these systematic errors at small gap heights. For example, a simple linear approximation has been proposed in which a simple gap error is defined to correct the measured values \citep{dudgeon1994domain, davies2008thin, connelly1985high}. Another technique involves using ultrasound time-of-flight measurements to detect the varying thickness of the fluid layer in the case of misalignment, which can be used to calculate the degree of misalignment \citep{rodriguez2013using}. A numerical solution of the flow in a misaligned parallel-plate rheometer was also presented by \citet{andablo2010method}. Also, \citet{clasen2013self} introduced a system that can self correct non-parallelism to a degree using hydrodynamic lubrication forces. Finally, a theoretical description of the velocity and stress profiles in a slightly misaligned cone-and-plate rheometer was achieved by \citet{dudgeon1994domain} using a domain perturbation study in the limit of zero Reynolds number.

In this section, we consider the role of misalignment on the transport of absorbed water throughout the glycerol, and the effects of this misalignment on the measured viscosity values. In general, such an analysis would need fully three-dimensional simulations in a misaligned rheometer geometry. We considered performing such simulations, but found them to be intractable due to the extremely high computational cost of performing them. In particular, they require 2-3 orders of magnitude more grid cells than the axisymmetric simulations in order to resolve the water concentration boundary layer at the outer edge of the rheometer. Furthermore, all of the meshing techniques we tried that would maintain this resolution at the outer edge ultimately resulted in very high aspect ratio cells at some point in the domain that affect resolution and greatly increase the number of iterations needed to solve the velocity/pressure profile with the SIMPLE algorithm, which must be repeated continuously as the water concentration field evolves. For these reasons, we consider a simplified, depth-averaged case that is valid in the limit of small gap heights. This model and the corresponding simulations and results will be described in the following sections.

\subsection{Theory}

\begin{figure}
\centering\includegraphics[width=0.7\textwidth]{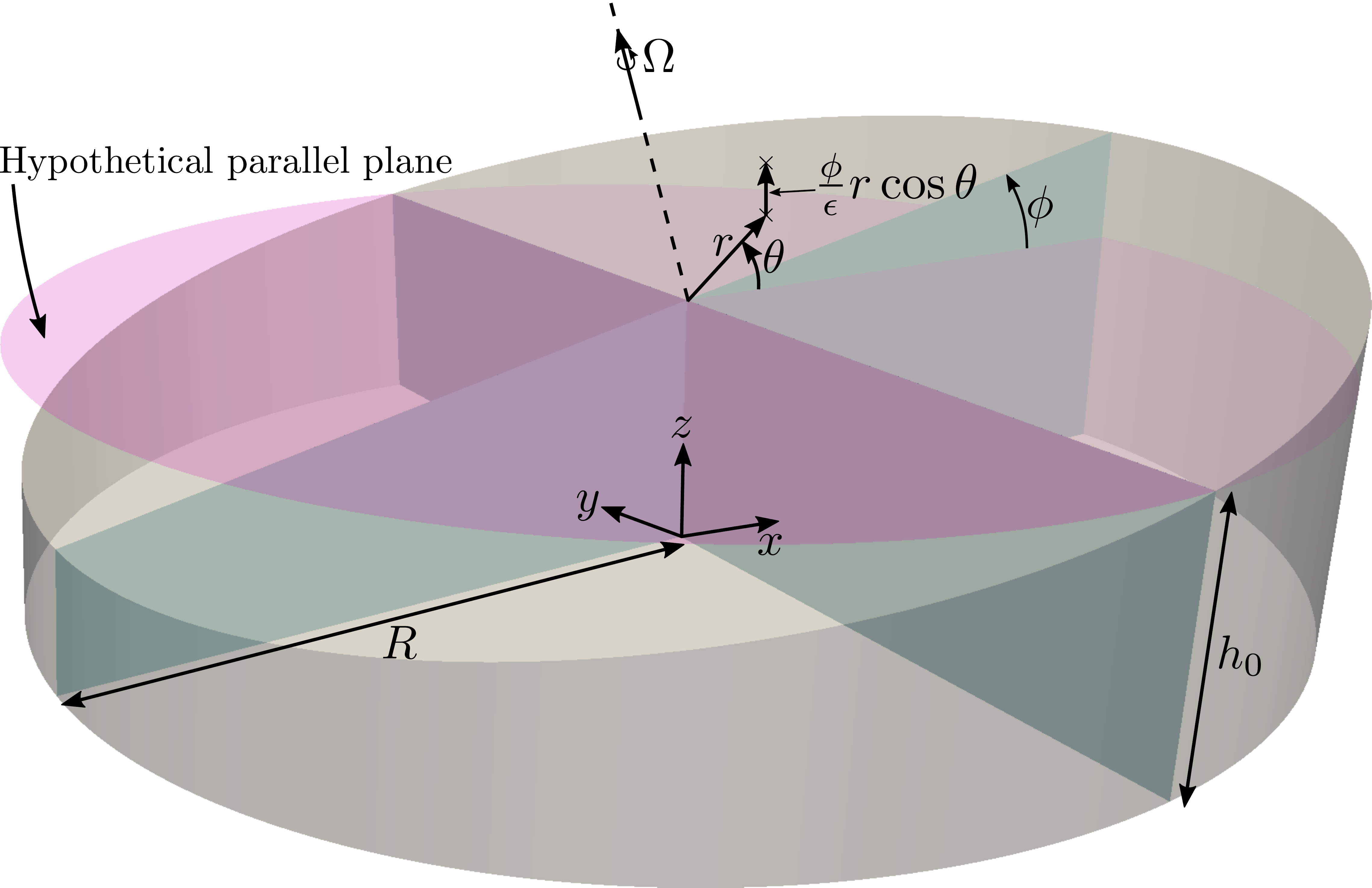}
\caption{Misaligned parallel-plate rheometer geometry and coordinate system. The upper plate is misaligned by a small deflection angle $\phi$ and rotates at angular speed $\Omega$. With $z$ nondimensionalized by $h_0$ and $r$ nondimensionalized by $R$, the $z$-coordinate defining the upper plate is $h(r,\theta,\phi)=1+\phi\epsilon^{-1}r\cos\theta$. Note that for small angles, the angle $\phi$ can range from 0 to a maximum of $\epsilon$.\label{fig:geometry}}
\end{figure}

Here, we consider a misaligned parallel-plate rheometer with radius $R$ and gap thickness $h(r,\theta,\phi)$, where the upper plate is slightly tilted by the small angle $\phi$. Once again, the lower plate is stationary, and the upper plate rotates at a rotation speed of $\Omega$. The coordinate system and problem setup are shown in Figure \ref{fig:geometry}. The boundary conditions for the system are $\boldsymbol{u}=\boldsymbol{0}$ at $z=0$ (on the lower plate), and 
\begin{subequations}\label{eq:BCs}
\begin{align}
u_r&=r\cos\theta\sin\theta\sin\phi\tan\phi,\\
u_\theta&=r\left(\cos^2\theta\sec\phi+\cos\phi\sin^2\theta\right),\\
u_z&=-r\epsilon^{-1}\sin\theta\sin\phi,
\end{align}
\end{subequations}
at $z= h(r,\theta,\phi)=1+\phi\epsilon^{-1}r\cos\theta$ (the upper plate). Note that these simplify to $u_r=u_z=0$ and $u_\theta=r$ at $z=1$ as in the axisymmetric case when $\phi=0$. For small angles, the angle $\phi$ can range from 0 to a maximum of $\epsilon$. Thus, $\phi/\epsilon$ ranges from 0 to 1, and small values of $\phi/\epsilon$ correspond to small plate deflections. Here, $\phi/\epsilon=0$ corresponds to the case of no misalignment, and $\phi/\epsilon=1$ corresponds to the case where the plates come in contact at one edge. Further, recall that as before we generally also need to apply boundary conditions at $r=1$. In a practical experiment, this boundary condition represents a fluid-air interface that is typically not flat and experiences surface tension effects. However, in the small gap limit, we lose the ability to impose such a boundary condition, and we note that this contributes to the error in velocity/pressure profiles in the $\mathcal{O}(\epsilon)$ region near $r=1$.

The governing equations are again the Navier-Stokes equations with variable viscosity and the continuity equation, which are given by Eqs. (\ref{eq:variableNS_cylindrical}) and (\ref{eq:continuity}), respectively, as well as the water concentration advection--diffusion equation given by Eq. (\ref{eq:solute_nondim}). As mentioned above, the numerical simulation of the full system of coupled equations in a well-resolved 3D geometry is computationally expensive. In the limit of narrow gap heights $\epsilon\ll1$ and negligible inertia $\Rey\ll1$, the Navier-Stokes equations simplify to
\begin{subequations}
\begin{align}
&0=-\frac{\partial p}{\partial r} +\mu\frac{\partial^2u_r}{\partial {z}^2} + \frac{\partial \mu}{\partial z}\frac{\partial u_r}{\partial z},\\
&0=-\frac{1}{r}\frac{\partial p}{\partial\theta}+\mu\frac{\partial^2u_\theta}{\partial {z}^2} + \frac{\partial \mu}{\partial z}\frac{\partial u_\theta}{\partial z},\\
&0=\frac{\partial p}{\partial z}.
\end{align}
\end{subequations}
Furthermore, in the thin gap limit, the water concentration can be assumed to be approximately uniform in the depth direction, which gives $\frac{\partial c}{\partial z}\approx0$ and $\frac{\partial\mu}{\partial z}\approx0$. This gives
\begin{equation}\label{eq:tilted_equations}
0=-\frac{\partial p}{\partial r} +\mu\frac{\partial^2u_r}{\partial {z}^2},~~~0=-\frac{1}{r}\frac{\partial p}{\partial\theta}+\mu\frac{\partial^2u_\theta}{\partial {z}^2},~~~0=\frac{\partial p}{\partial z},~~~\text{and}~~~\frac{1}{r}\frac{\partial}{\partial r}\left(r u_r\right) + \frac{1}{r}\frac{\partial u_\theta}{\partial \theta} + \frac{\partial u_z}{\partial z} = 0.
\end{equation}
Note here that $c(r,\theta,t)$, $\mu(r,\theta,t)$ and $p(r,\theta,t)$ in this limit. With the fact that $\partial p/\partial z=0$ in this limit, the next leading-order form of the $z$-component of the Navier-Stokes equations becomes
\begin{equation}\label{eq:next_leading_order}
0=\mu\frac{\partial^2 u_z}{\partial z^2}+\frac{\partial\mu}{\partial r}\frac{\partial u_r}{\partial z} + \frac{1}{r}\frac{\partial \mu}{\partial\theta}\frac{\partial u_\theta}{\partial z}.
\end{equation}

By examining the form of the gap height distribution $h(r,\theta,\phi)=1+\phi\epsilon^{-1}r\cos\theta$ we see that the magnitude of the perturbation is $\mathcal{O}(\phi/\epsilon)$. This suggests the use of a solution given by
\begin{subequations}\label{eq:expansion}
\begin{gather}
u_r(r,\theta,z,t)=\left(\frac{\phi}{\epsilon}\right)u_{r,1}(r,\theta,z,t)+\left(\frac{\phi}{\epsilon}\right)^2u_{r,2}(r,\theta,z,t)+\hdots,\\
u_\theta(r,\theta,z,t)=rz+\left(\frac{\phi}{\epsilon}\right)u_{\theta,1}(r,\theta,z,t)+\left(\frac{\phi}{\epsilon}\right)^2u_{\theta,2}(r,\theta,z,t)+\hdots,\\
u_z(r,\theta,z,t)=\left(\frac{\phi}{\epsilon}\right)u_{z,1}(r,\theta,z,t)+\left(\frac{\phi}{\epsilon}\right)^2u_{z,2}(r,\theta,z,t)+\hdots,\\
p(r,\theta,t)=\left(\frac{\phi}{\epsilon}\right)p_1(r,\theta,t)+\left(\frac{\phi}{\epsilon}\right)^2p_2(r,\theta,t)+\hdots,
\end{gather}
\end{subequations}
which is valid in the limit $\phi/\epsilon\ll1$. Substituting this expansion into Eqs. (\ref{eq:tilted_equations}) and (\ref{eq:next_leading_order}) and applying the boundary conditions gives
\begin{subequations}\label{eq:tilted_velocities}
\begin{align}
u_{r,1}(r,\theta,z,t)=&\frac{z(z-1)}{2\mu}\frac{\partial p_1}{\partial r},\\
u_{\theta,1}(r,\theta,z,t)=&-r^2 z \cos\theta + \frac{z(z-1)}{2 r \mu}\frac{\partial p_1}{\partial\theta},\\
\nonumber u_{z,1}(r,\theta,z,t)=&-\frac{z^2}{12r^2\mu}\left\{6r^3\mu^2\sin\theta - (2z-3)\left[\frac{\partial\mu}{\partial\theta}\frac{\partial p_1}{\partial\theta} + r^2\frac{\partial\mu}{\partial r}\frac{\partial p_1}{\partial r}\right]\right.\\
&\left. + (2z-3)\mu\left[\frac{\partial^2 p_1}{\partial \theta^2}+r\left(\frac{\partial p_1}{\partial r}+r\frac{\partial^2 p_1}{\partial r^2}\right)\right]\right\}.
\end{align}
\end{subequations}
This procedure also yields a PDE governing the pressure distribution that is given by
\begin{equation}\label{eq:pressure}
\mu\left\{\frac{\partial^2 p_1}{\partial\theta^2} + r\left[\frac{\partial p_1}{\partial r} + r\left(6r\mu\sin\theta + \frac{\partial^2 p_1}{\partial r^2}\right)\right]\right\} = \frac{\partial\mu}{\partial\theta}\frac{\partial p_1}{\partial\theta} + r^2\frac{\partial\mu}{\partial r}\frac{\partial p_1}{\partial r}.
\end{equation}
With some known distribution of viscosity in the system, a numerical solution of Eq. (\ref{eq:pressure}) yields the pressure distribution in the gap. This in turn can be used to calculate the velocity profiles from Eqs. (\ref{eq:expansion}) and (\ref{eq:tilted_velocities}). The velocity profiles can then be used to update the water concentration distribution via the advection--diffusion equation. With the assumption that $c$ is independent of $z$ (valid in the small gap limit), the solute transport equation becomes
\begin{equation}\label{eq:solute_nondim_small_gap1}
\Pe\frac{\partial c}{\partial t}=\frac{1}{r}\frac{\partial}{\partial r}\left(rD\frac{\partial c}{\partial r}\right)+\frac{1}{r^2}\frac{\partial}{\partial\theta}\left(D\frac{\partial c}{\partial\theta}\right)-\Pe\left(u_r\frac{\partial c}{\partial r}+\frac{u_\theta}{r}\frac{\partial c}{\partial \theta}\right).
\end{equation}
Since $c$ is independent of $z$, we consider solving the depth-averaged version of this equation instead, which is simply
\begin{equation}\label{eq:solute_nondim_small_gap2}
\Pe\frac{\partial c}{\partial t}=\frac{1}{r}\frac{\partial}{\partial r}\left(rD\frac{\partial c}{\partial r}\right)+\frac{1}{r^2}\frac{\partial}{\partial\theta}\left(D\frac{\partial c}{\partial\theta}\right)-\Pe\left(\bar u_r\frac{\partial c}{\partial r}+\frac{\bar u_\theta}{r}\frac{\partial c}{\partial \theta}\right),
\end{equation}
where bars denote depth-averaged quantities, $\bar c = c$, $\bar D = D$, and
\begin{subequations}\label{eq:depth_averaged}
\begin{gather}
\bar u_r = \frac{1}{1+\frac{\phi}{\epsilon}r\cos\theta}\int\limits_{z=0}^{1+\frac{\phi}{\epsilon}r\cos\theta} u_r(r,\theta,z) \mathrm{d}z = -\frac{1}{12\mu}\frac{\phi}{\epsilon}\frac{\partial p_1}{\partial r} + \mathcal{O}\left(\frac{\phi}{\epsilon}\right)^2,\\
\bar u_\theta = \frac{1}{1+\frac{\phi}{\epsilon}r\cos\theta}\int\limits_{z=0}^{1+\frac{\phi}{\epsilon}r\cos\theta} u_\theta(r,\theta,z) \mathrm{d}z = \frac{r}{2} - \frac{1}{12 r \mu}\frac{\phi}{\epsilon}\frac{\partial p_1}{\partial\theta} + \mathcal{O}\left(\frac{\phi}{\epsilon}\right)^2.
\end{gather}
\end{subequations}
Thus, having solved the pressure distribution due to the misalignment $p_1$ from Eq. (\ref{eq:pressure}), the depth-averaged velocity components can be calculated from Eq. (\ref{eq:depth_averaged}), which can in turn be used to advect the solute concentration.

\subsection{Numerical methods}

We perform numerical simulations of the coupled transport equations described in the previous section. Recall that we seek solutions of the water transport and associated viscosity measurements in a misaligned parallel-plate rheometer that is valid at small gap heights and in the limit of negligible inertia. The numerical approach for solving these systems is as follows:
\begin{enumerate}
\item Solve Eq. (\ref{eq:pressure}) for the pressure perturbation due to misalignment subject to the boundary condition $p\rightarrow0$ at $r=1$.
\item Calculate the depth-averaged radial and azimuthal velocity components from Eq. (\ref{eq:depth_averaged}).
\item Advance the water concentration profile in time by numerically integrating Eq. (\ref{eq:solute_nondim_small_gap2}) for one or more timesteps.
\item Calculate the new viscosity and diffusivity fields and iterate back to Step 1.
\end{enumerate}
We attempt a numerical implementation of this process using a finite-difference implementation in MATLAB. However, several complications emerge due to the extremely large Peclet numbers in the system. In particular, the diffusion of the water concentration field is so slow that the field effectively propagates with a very sharp front. In order to resolve this and avoid spurious oscillations in the concentration field, we use slope-limited finite differencing based on the minmod limiter function to switch to first-order spatial differencing at the steep gradient \citep{roe1986characteristic}. This avoids the oscillations that result in a pure second-order differencing scheme and still allows nearly second-order accuracy in space globally. The more serious difficulty that we encountered with this numerical approach is solving Eq. (\ref{eq:pressure}) for the perturbation pressure field. These solutions do not behave nicely due to the sharp viscosity gradients on the right-hand side of the equation. We were not able to resolve this issue using slope-limiters.

As an alternative approach and to illustrate the qualitative dynamics that can be expected with a misaligned upper plate, we instead assume a constant viscosity model for the purposes of calculating the velocity profile, since this solution is well-behaved. Note that Eq. (\ref{eq:pressure}) has an analytical solution when $\mu=1$ which is given by
\begin{equation}
p_1(r,\theta) = -\frac{3}{4}r\left(-1+r^2\right)\sin\theta.
\end{equation}
We use this theoretical result at constant viscosity to calculate the depth-averaged velocities in the misaligned rheometer and use these to update the concentration profile. When calculating the effective measured viscosity and torque, we always use the viscosity distribution that corresponds to the concentration profile. This limitation to a velocity profile based on constant viscosity is a clear limitation of our results, but nevertheless they capture qualitative features of the experiments that the axisymmetric model could not predict, and we leave a full solution with evolving velocity profiles based on spatial variation of viscosity to future work.

\subsection{Results}

\begin{figure}
\centering\includegraphics[width=0.8\textwidth]{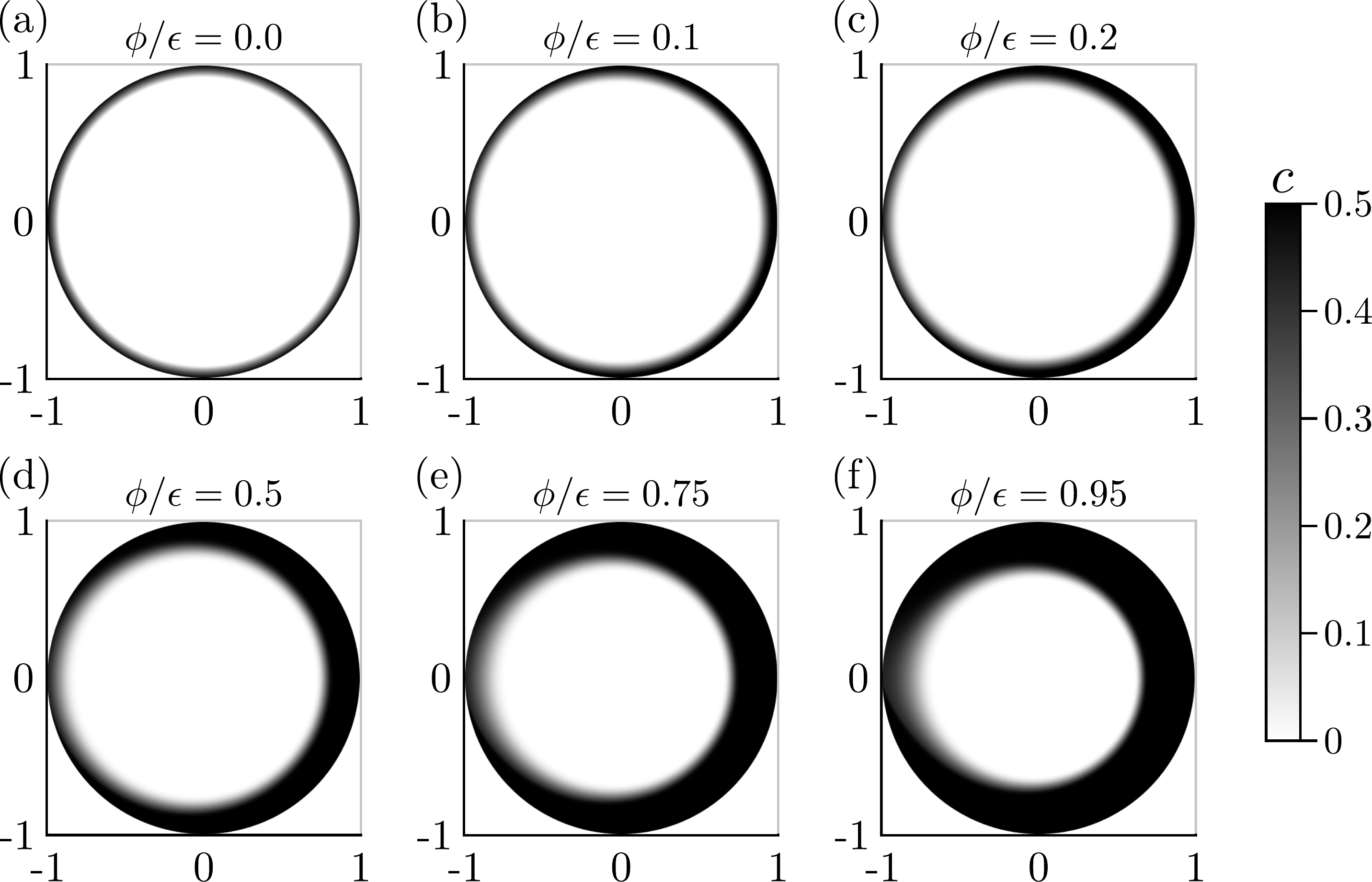}
\caption{Depth-averaged water concentration profile at $t^*=3600$ s as a function of misalignment for $c_\text{sat}=0.5$ and $\Omega=0.4$ rad/s. Results correspond to $\phi/\epsilon$ values of (a) 0.0 (perfectly aligned), (b) 0.1, (c) 0.2, (d) 0.5, (e) 0.75, and (f) 0.95 (plates are nearly contacting). As can be seen, as the misalignment increases, the concentration profile becomes no longer axisymmetric, and there is a significant increase in total water transport into the fluid layer from the outer edge due to the misalignment-driven secondary flows.\label{fig:tilting}}
\end{figure}

Here, we introduce the numerical results achieved for the misaligned rheometer system based on the methodology described in the previous section. First, we highlight the role of the misalignment on the water concentration field in Figure \ref{fig:tilting}.
Here, the depth-averaged water concentration profiles are shown at $t^*=3600$ s as a function of misalignment for $c_\text{sat}=0.5$ with misalignment ranging from $\phi/\epsilon=0$ to 0.95, where 0 is the perfectly aligned parallel-plate case, and 1 is the limit where the plates come into contact at one edge. As can be seen, the misaligned cases all show a non-axisymmetric concentration profile. This is due to the non-axisymmetric secondary velocity components due to the misalignment. In particular, the radial component $\bar u_r$ transports water towards or away from the outer edge, and is $\bar u_r \sim \frac{\phi}{\epsilon}\frac{\partial p_1}{\partial r}$. With $p_1\sim \sin\theta$, this represents a radially inward flow on one half of the rheometer and a radially outward flow on the other half and explains why in Figure \ref{fig:tilting} the concentration profile appears to be pulled in from the right edge and pushed towards the left edge. Furthermore, this secondary velocity component is proportional to $\phi/\epsilon$, so doubling the degree of misalignment doubles the radial advective fluxes in both directions.

\begin{figure}
\centering\includegraphics[width=0.5\textwidth]{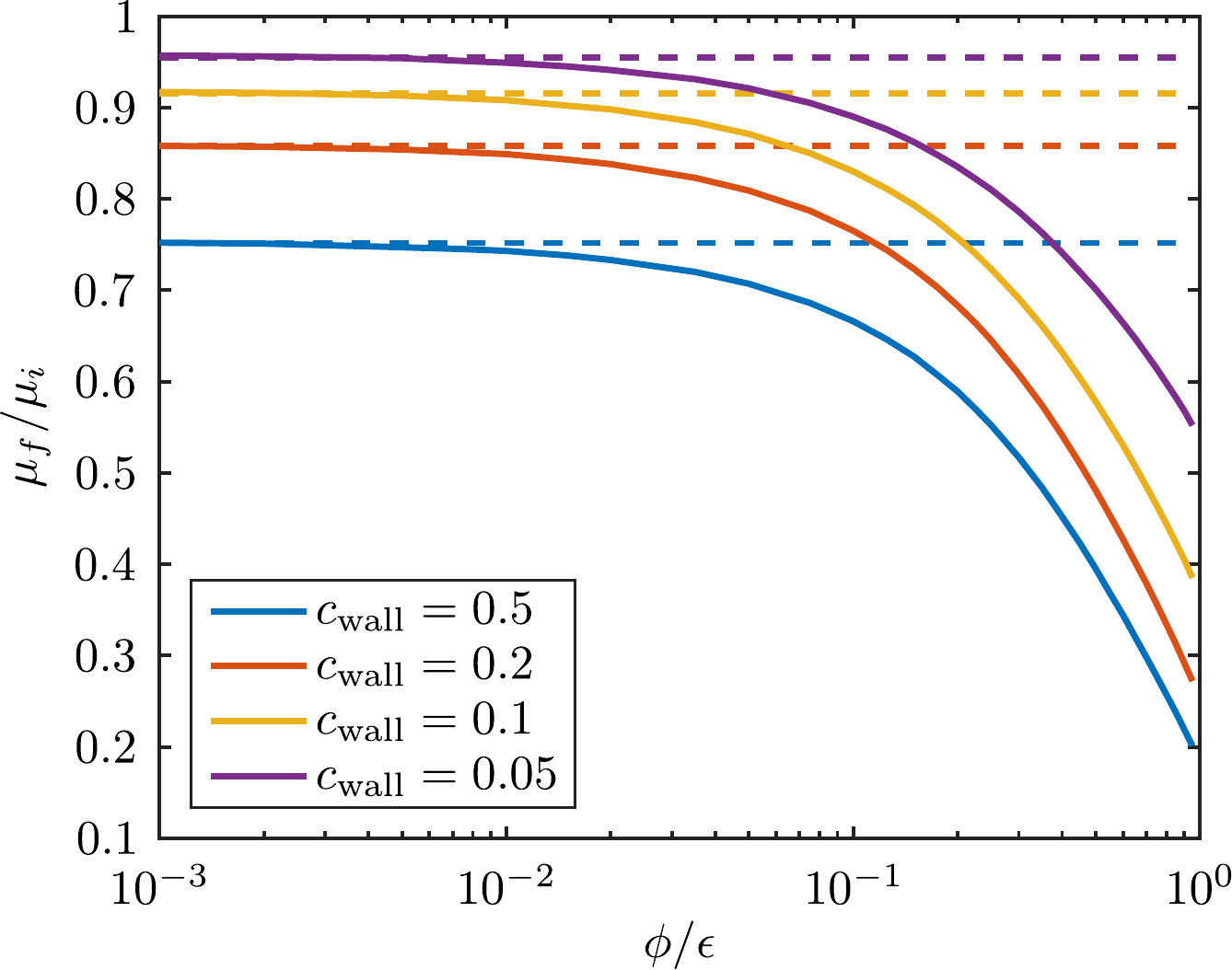}
\caption{Final measured viscosity values $\mu_f/\mu_i$ at $t^*=3600$ s as functions of the misalignment $\phi/\epsilon$ and $c_\text{sat}$ at an angular rotation speed of 0.4 rad/s. Dashed lines correspond to the 1D pure-diffusion limit. Results asymptotically approach the 1D diffusion-dominated limit as $\phi/\epsilon\rightarrow0$. Furthermore, results show a steep drop-off in measured viscosity values as the misalignment increases.\label{fig:tilted_results}}
\end{figure}

In order to quantify the effect of misalignment on the measured viscosity values, simulations were performed across a range of $\phi/\epsilon$ and $c_\text{sat}$ values. The final measured viscosity values from these simulations are presented in Figure~\ref{fig:tilted_results}.
Here, the results correspond to an angular speed of 0.4 rad/s. The dashed lines in the figure correspond to the 1D diffusion-dominated regime, and the results asymptotically approach these limits as $\phi/\epsilon\rightarrow0$. Furthermore, the results show a steep dropoff in measured final viscosities at large misalignments, which possibly explains the sharp decrease in measured viscosity in the experiments at small gap heights (e.g., Figure \ref{fig:experimental}b) that was not captured in the axisymmetric model (e.g., Figure \ref{fig:axisymmetric_compiled}).

\begin{figure}
\centering\includegraphics[width=\textwidth]{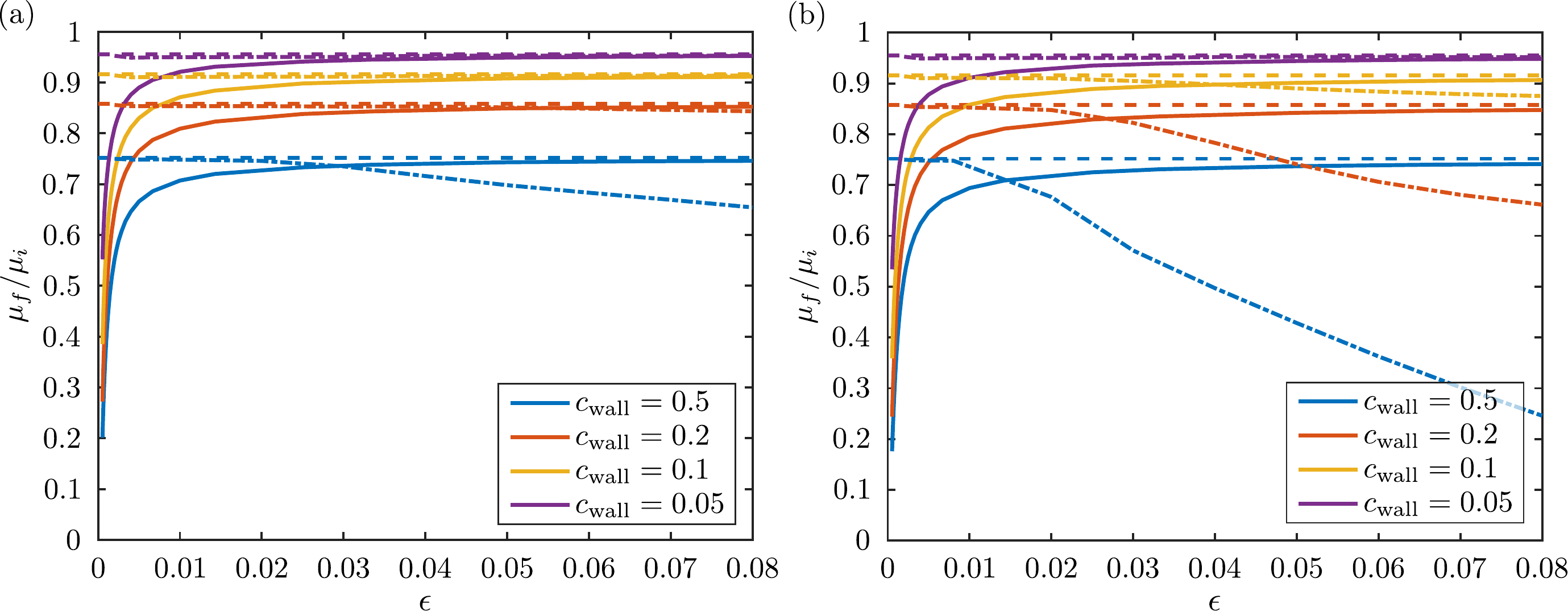}
\caption{Comparison of the final measured viscosity values at $t^*=3600$ s for the models corresponding to each of the three regimes: (1) the 1D axisymmetric, pure-diffusion limit (dashed lines), (2) the axisymmetric, inertial regime (dot-dashed lines), and (3) the misaligned, inertialess, small gap limit (solid lines). Results correspond to angular rotation speeds of (a) 0.4 rad/s and (b) 1.0 rad/s. Misaligned cases are calculated with a misalignment angle of 0.0005 rad. All cases were performed with a rheometer of radius $R=2.5$ cm.\label{fig:tilted_vs_axisymmetric}}
\end{figure}

Finally, a comparison between all of the three different proposed models (i.e., the 1D pure diffusion limit, the axisymmetric inertial limit, and the misaligned inertialess small gap limit) is shown in Figure \ref{fig:tilted_vs_axisymmetric}.
Here, the dashed lines indicate the 1D diffusion-dominated limit, dot-dashed lines correspond to the inertial, axisymmetric regime, and the solid lines show the results for the misaligned, inertialess, small-gap limit. Results were calculated for angular rotation speeds of (a) 0.4 rad/s and (b) 1.0 rad/s. The results show that at small gap heights, the misalignment effects dominate the inertial effects. This becomes more clear when considering that the radial secondary velocity due to inertial effects in an axisymmetric case is $\mathcal{O}(\Rey)$, whereas the secondary velocity components due to misalignment are $\mathcal{O}(\phi/\epsilon)$. For fixed angular rotation speed, the secondary velocities must dominate the inertial secondary velocities as the gap height decreases. Furthermore, the results also show that the opposite is true at large gap heights. For a fixed misalignment angle and rotation speed, $\phi/\epsilon$ decreases as the gap height increases, whereas the inertial effects increase, such that the large gap regime is dominated by inertial effects. This cross-over explains the non-monotonic relationship between $\mu_f$ and gap height reported in Figure \ref{fig:experimental}b. Thus, there is a critical $\epsilon$ value at which the measured viscosity values switch from being misalignment-dominated to inertia-dominated. Numerical simulations using the previously described models and numerical methods can be used to estimate this transition, as shown in Figure \ref{fig:tilted_vs_axisymmetric}.

\section{Conclusions}

In this paper, we have considered the measurement of the viscosity of glycerol in a parallel-plate rheometer. Intuitively, it can be anticipated that the viscosity must decrease over time due to the hygroscopic nature of the fluid as it absorbs water vapor from the atmosphere. Based on an initial understanding of the fluid dynamics in a parallel-plate rheometer for a constant viscosity flow, an axisymmetric model of the flow predicts that the dynamics should be purely limited by diffusion in the thin-gap limit and become independent of gap height. However, a sharp drop-off in measured viscosity values was observed experimentally at small gap heights, which motivated us to reconsider the fluid dynamics in the system and led to the hypothesis that plate misalignment could drive additional secondary flows that might affect the transport of the water concentration throughout the system. Ultimately, theoretical models and numerical simulations of the coupled dynamics and measured viscosity values were achieved in three different regimes: (1) the 1D inertialess, diffusion-dominated regime, (2) the axisymmetric inertial regime, and (3) the misaligned, inertialess, thin-gap regime. Results confirmed that there are two types of secondary flows that can exist in such systems. The first of these is $\mathcal{O}(\Rey)$ and corresponds to the secondary inertially driven flows in a perfectly aligned axisymmetric parallel-plate rheometer. The other secondary flow is $\mathcal{O}(\phi/\epsilon)$ and is driven by the $\mathcal{O}(\phi/\epsilon)$ plate misalignment. Assuming a fixed misalignment $\phi$, then as the gap height decreases, the $\mathcal{O}(\phi/\epsilon)$ misalignment flow will inevitably dominate the $\mathcal{O}(\Rey)$ inertially driven flow.

\begin{figure}
\centering\includegraphics[width=0.6\textwidth]{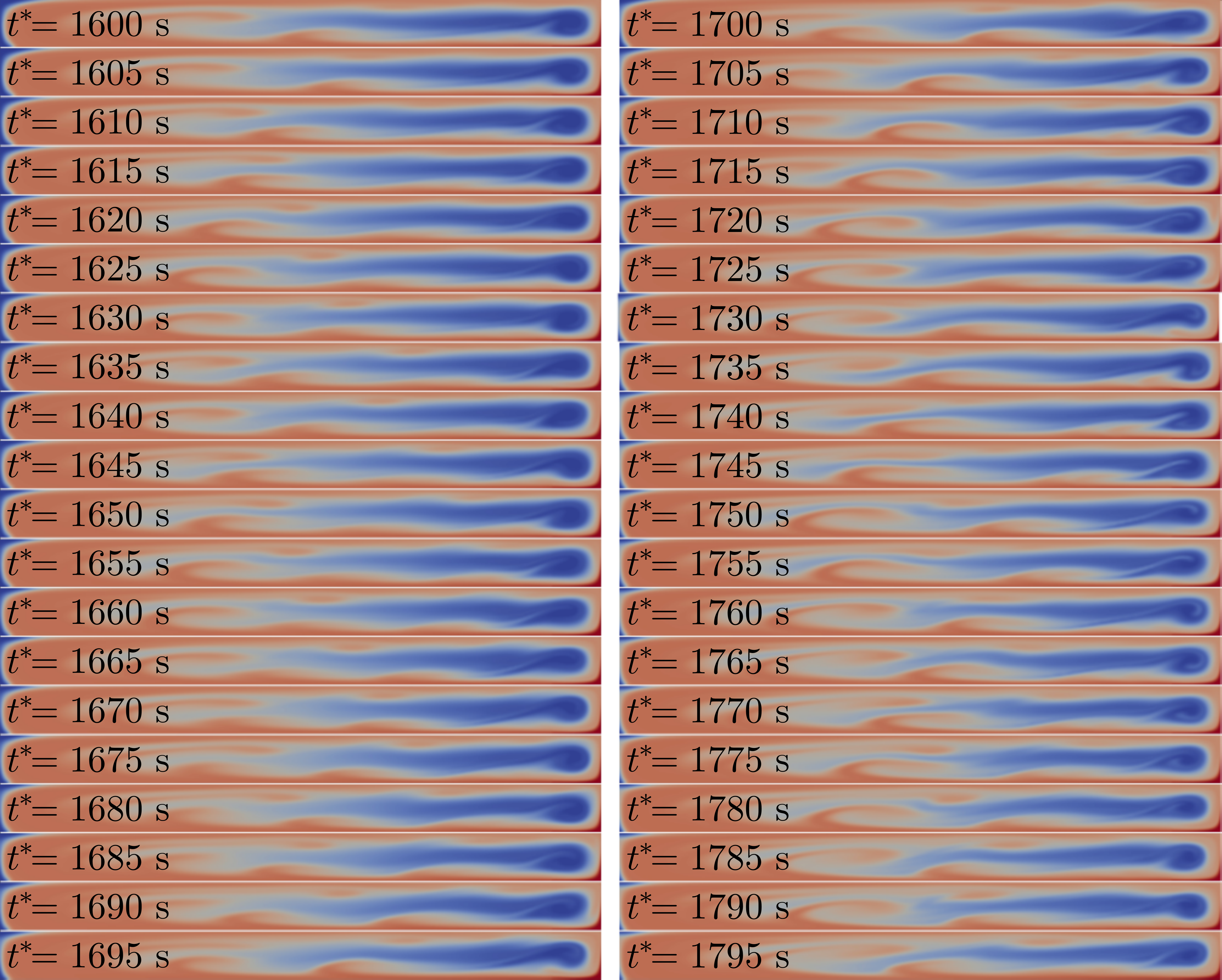}
\caption{Apparent instability/oscillation in the concentration profile field due to viscosity gradients from numerical results with $\Omega=10$ rad/s, $c_\text{sat}=0.2$, and $h_0=2$ mm.\label{fig:oscillation}}
\end{figure}

Based on the results here via comparison between experiments and numerical simulations, as well as between simulations with parallel and misaligned plates, we argue that the sharp decrease in measured viscosity is attributable to the secondary flows induced by plate misalignment. The mechanism by which the misalignment results in a faster decrease in viscosity seems to be that the secondary velocities pull the relatively high water concentration away from the outer boundary, which steepens the concentration gradient at the outer boundary, resulting in an increased mass flux of water into the glycerol, which subsequently lowers the viscosity of the glycerol. These results have relevance not only to the measurement of the viscosity of glycerol solutions (for which care must be taken to ensure all the possible transport mechanisms are understood), but also for our understanding of the flow in parallel-plate rheometers more generally. We have shown that misalignment effects in particular can have a surprising critical influence over the mass transport in such systems, especially as the gap height becomes small. Furthermore, based on these results, it is plausible that such viscosity measurements of glycerol in a parallel-plate rheometer could potentially be used as a technique to quantify the degree of misalignment in the rheometer, although we leave a practical investigation and demonstration of this technique for future work. Finally, we note that additional complications can arise in such a system, such as the potential for the emergence of instabilities due to viscosity gradients. We observed evidence for such effects in our numerical results at high angular rotation speeds in some cases. For example, Figure \ref{fig:oscillation}
shows a time series of the concentration profile for a case with $\Omega=10$ rad/s, $c_\text{sat}=0.2$, and $h_0=2$ mm. It is known that viscosity stratification in shear flows can lead to  instability in certain regimes \citep{yih1967instability, sahu2014instability}. However, the coupled viscosity distribution and velocity profile in the rheometer system are highly nonlinear, and the flow cannot be analyzed in terms of simple viscosity-stratified layers of fluid. Furthermore, at the high rotation speeds and gap heights where we observed this instability, it is likely that other assumptions in our proposed models will break down, especially the assumed flat interface at the outer boundary and the predefined slip boundary conditions there. Thus, we leave a detailed study of these intriguing instabilities for future work.

Declaration of interests: The authors report no conflict of interest.

Author contributions: J.T.A. developed the theory and performed the numerical simulations. A.P., A.G., and S.S. performed the experiments. J.T.A., S.S., and H.A.S. conceived the experiment and wrote the manuscript. The authors would like to acknowledge Ian Jacobi for helpful discussions regarding the experimental setups.

\bibliography{references}

\end{document}